\documentclass[iop,numberedappendix]{emulateapj}
\usepackage{epsfig}
\usepackage{amsmath}
\usepackage{enumerate}
\usepackage{natbib}
\usepackage{hyperref}

\usepackage{ulem}
\usepackage[squaren, Gray, cdot]{SIunits}
\usepackage{color}

\def\jpm{j_{\pm}}
\def\kpm{\kappa_{\pm}}

\def\dotg{\dot{\gamma}}
\def\dotp{\dot{p}}
\def\tauT{\tau_{\rm T}}
\def\epsB{\varepsilon_{B}}
\def\epsh{\varepsilon_{\rm h}}
\def\rcoll{R_{\rm c}}
\def\rph{R_\star}

\def\Lh{L_{\rm h}}
\def\Lpl{L_{\rm pl}}
\def\Lrad{L_{\rm rad}}
\def\LB{L_{B}}
\def\Lturb{L_{\rm turb}}

\def\sigmaT{\sigma_{\rm T}}
\def\afs{\alpha_{\rm f}}
\def\me{m_{\rm e}}
\def\mprot{m_{\rm p}}
\def\Ne{n_{\rm e}}
\def\EB{E_{\rm B}}

\def\kB{k_{\rm B}}
\def\tauT{\tau_{\rm T}}
\def\taup{\tau_{\rm p}}
\def\taupm{\tau_{\pm}}

\def\Te{T_{\rm e}}
\def\TB{T_{\rm B}}

\def\sigmat{\sigma_{\rm T}}
\def\ginj{\gamma_{\rm inj}}
\def\pinj{p_{\rm inj}}
\def\Qinj{Q_{\rm inj}}
\def\Urad{U_{\rm rad}}
\def\UB{U_{\rm B}}
\def\epsrad{\varepsilon_{\rm rad}}
\def\epsB{\varepsilon_{\rm B}}
\def\epsinj{\varepsilon_{\rm nth}}

\def\epsth{\varepsilon_{\rm th}}
\def\epsnth{\varepsilon_{\rm nth}}
\def\Lth{L_{\rm th}}
\def\Lnth{L_{\rm nth}}
\def\kh{k_{\rm h}}
\def\kth{k_{\rm th}}
\def\knth{k_{\rm nth}}
\def\yC{y_{\rm C}}

\def\Lrad{L_{\rm rad}}
\def\dotNsynch{\dot{\cal{N}}_{\rm synch}}
\def\alphacasc{\alpha_{\pm}}
\def\gcascmin{\gamma_{\rm \pm,min}}
\def\Ms{\cal{M}_{\rm s}}
\def\dotNph{\dot{N}_{\rm ph}}

\def\Qinj{Q_{\rm inj}}

\def\Inu{I_{\nu}}
\def\jnu{j_{\nu}}
\def\nocc{\tilde{n}}
\def\jocc{\tilde{j}_{\nu}}
\def\jav{\overline{j}}
\def\knu{\kappa_{\nu}}
\def\d{\partial}
\def\dotM{\dot{M}}
\def\Zpm{Z_{\pm}}

\def\npm{n_{\pm}}
\def\nprot{n_{\rm p}}
\def\epm{e^{\pm}}
\def\rW{R_{\rm W}}
\def\thetaC{\theta_{\rm C}}
\def\thetae{\theta_{\rm e}}

\def\ninj{\dot{n}_{\pm}}
\def\nann{\dot{n}_{\rm ann}}

\def\enthinit{\varepsilon_{0,{\rm nth}}}
\def\ethinit{\varepsilon_{0,{\rm th}}}
\def\epshinit{\varepsilon_{0,{\rm h}}}

\def\Epk{E_{\rm pk}}
\def\Rph{R_\star}
\def\nph{n_{\rm ph}}
\def\Eq{Equation}

\newbox\grsign \setbox\grsign=\hbox{$>$} \newdimen\grdimen \grdimen=\ht\grsign
\newbox\simlessbox \newbox\simgreatbox \newbox\simpropbox
\setbox\simgreatbox=\hbox{\raise.5ex\hbox{$>$}\llap
     {\lower.5ex\hbox{$\sim$}}}\ht1=\grdimen\dp1=0pt
\setbox\simlessbox=\hbox{\raise.5ex\hbox{$<$}\llap
     {\lower.5ex\hbox{$\sim$}}}\ht2=\grdimen\dp2=0pt
\setbox\simpropbox=\hbox{\raise.5ex\hbox{$\propto$}\llap
     {\lower.5ex\hbox{$\sim$}}}\ht2=\grdimen\dp2=0pt
\def\simgt{\mathrel{\copy\simgreatbox}}
\def\simlt{\mathrel{\copy\simlessbox}}

\shorttitle{Radiative transfer models for gamma-ray bursts}
\shortauthors{Vurm \& Beloborodov}

\begin{document}

\title{Radiative transfer models for gamma-ray bursts} 
 
\author{
Indrek Vurm\altaffilmark{1,2} and Andrei M. Beloborodov\altaffilmark{1}
}
\affil{$^1$Physics Department and Columbia Astrophysics Laboratory, Columbia University, 538 West 120th Street, New York, NY 10027, USA \\
$^2$Tartu Observatory, T\~{o}ravere 61602, Tartumaa, Estonia \\
}

%\maketitle
\label{firstpage}
\begin{abstract}
We present global radiative transfer models for heated relativistic jets.
The simulations include all relevant radiative processes, starting deep in the opaque zone 
and following the evolution of radiation to and beyond the photosphere of the jet. 
The transfer models are compared with three gamma-ray bursts 
GRB~990123, GRB~090902B, and GRB~130427A, which have well-measured 
and different spectra. 
The models provide good fits to the observed
spectra in all three cases, and we obtain estimates for the 
jet magnetization parameter $\epsB$ and the Lorentz factor $\Gamma$.
In the small sample of three bursts, $\epsB$ varies between 
0.01 and 0.05, and $\Gamma$ varies between 400 and 1200.

\end{abstract}


%############################################################

\section{Introduction}

Cosmological gamma-ray bursts (GRBs) have been observed for more than 
four decades, and thousands of bursts have rather well measured spectra. 
No physical model has been demonstrated to systematically
fit the observed spectra. Instead, the data are usually described using
a phenomenological Band function \citep{Band93} 
or a Band function combined 
with a blackbody and a power law
\citep{RydePeer09,Guiriec15}.
The spectrum peaks at photon energy $\Epk$ which is 
normally comparable to 1~MeV 
\citep{Kaneko06,Goldstein12} 
and almost never exceeds 10~MeV.

\subsection{Transparent or opaque source?}

A simple interpretation
of the observed nonthermal spectrum would be synchrotron emission from 
suddenly accelerated and quickly cooled electrons in an optically thin source.
Then the MeV peak of the GRB spectrum comes from the 
peak of the electron energy distribution
$\Ne(\gamma)$ and the gamma-ray tail comes
from the high-energy tail in $\Ne(\gamma)$.
This picture has to assume that the 
plasma has no dominant Maxwellian component.
Fitting the prompt data with synchrotron models using a Maxwellian plus power-law electron distribution
either requires the Maxwellian to be significantly weaker than the non-thermal component \citep{BaringBraby04,Burgess11}, or at most comparable in energy \citep{Burgess14}.
The peak of the GRB spectrum is then associated with a cutoff  
of the dominant nonthermal electron distribution at low energies.
The origin of such a special distribution is unknown. 
Simulations of possible heating mechanisms 
--- shocks and magnetic reconnection --- give different $\Ne(\gamma)$.
Shock heating results in a dominant Maxwellian component
with only a small fraction of particles populating the high-energy tail
\citep{SironiSpitkovsky2011},
and reconnection gives a broad flat distribution with 
no cutoff at low energies \citep{Kagan15}.

Regardless of the heating mechanism, the
synchrotron model faces a problem when it is compared with observations.
The synchrotron spectrum has a rather broad peak, even when the source is assumed to 
have a uniform magnetic field $B$ and the 
power-law electron distribution is assumed to
have a sharp cutoff at low energies. 
Attempts to reproduce GRB spectra with such 
idealized models gave acceptable fits for a small number of bursts 
\citep{BaringBraby04,Burgess11,Burgess14,Preece14}.
In a more realistic model, several factors inevitably broaden the synchrotron peak 
\citep[][hereafter B13]{Beloborodov13},
which makes it incompatible with observed spectra in virtually all 
GRBs \citep{Axelsson15,Yu2015}.

\begin{figure}[t]
\plotone{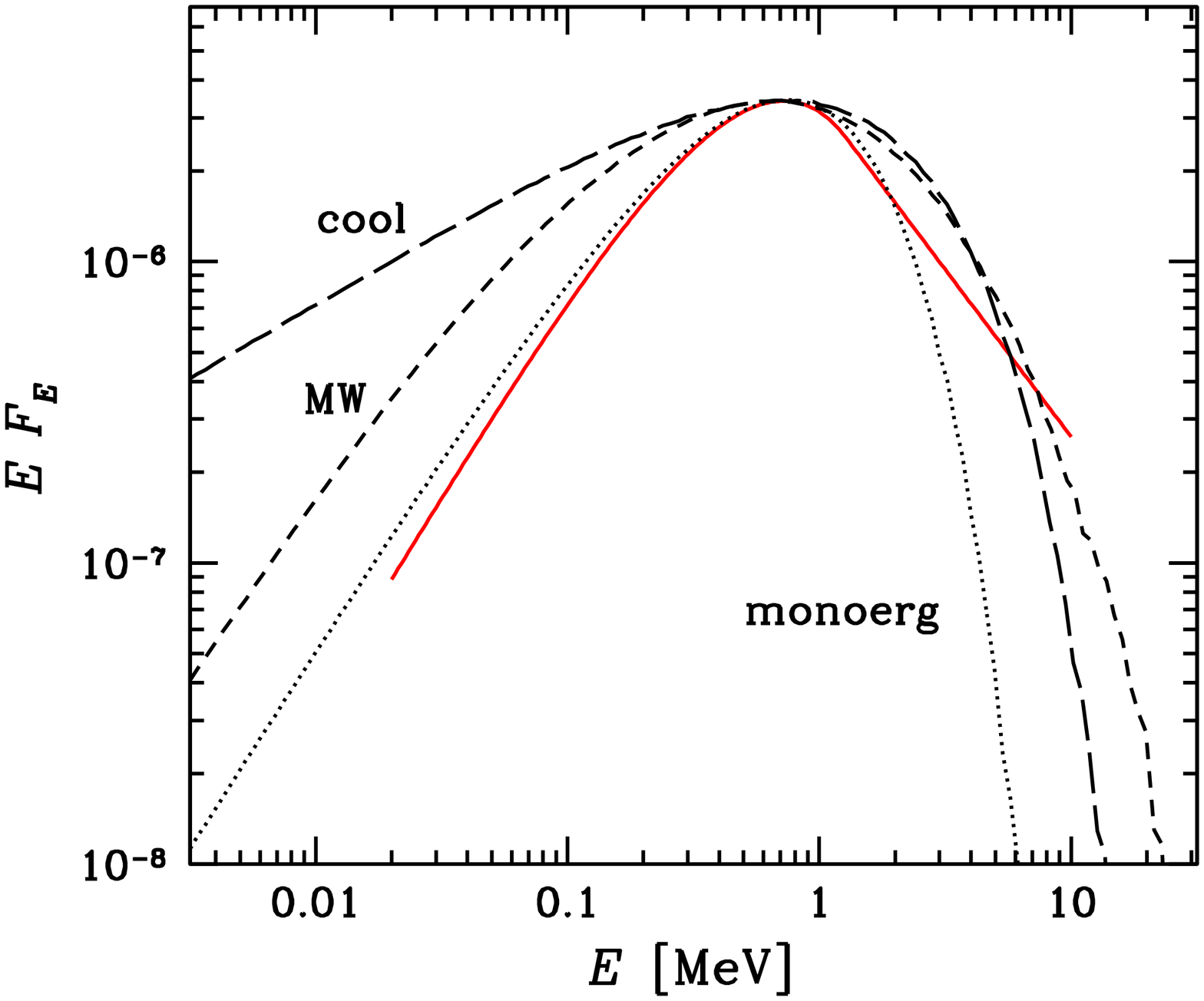}
\caption{Synchrotron spectra from an optically thin spherical shell
with three electron distributions: mono-energetic (dotted line), Maxwellian 
(short-dashed line), and fast-cooling Maxwellian
(long-dashed line). For comparison the Band fit of GRB 990123 
is shown by the red thick curve.
}
\label{fig:optthin}
\end{figure}

This problem is illustrated in Figure \ref{fig:optthin} where 
a synchrotron spectrum is compared with the Band fit of
a bright burst with a typical spectrum,  GRB~990123.
In the realistic fast-cooling regime, the minimum width of the synchrotron peak 
at half maximum exceeds $1.5$ orders in photon energy, 
even with a single value of $B$
in the emitting source and the narrow injected electron distribution (Maxwellian).
It is significantly broader than the Band fit to the {\it time-average} of the variable 
GRB spectrum \citep{Briggs99}.
Uncertainties in the observed spectrum due to the detector response and limited
photon statistics allow some room for stretching the measured peak width, however 
can hardly be made consistent with 
the synchrotron model, especially when the inevitable broadening due to variable 
magnetic field, electron injection, and Doppler factor is taken into account.

%%%%%%%%%%%%%%%%%%%%%%%%
%%%%%%%%%%%%%%%%%%%%%%%%

The sharp MeV spectral peak provides strong evidence for 
thermalization of radiation at early, opaque stages of the GRB explosion.
The inheritance of the spectral peak from an initial thermalization stage is also supported 
by the observed distribution of $\Epk$,
which cuts off above $\sim 10$~MeV
in agreement with a theoretical maximum (B13). 
In addition, many GRBs show hard spectral slopes violating
the ``line of death'' for optically thin synchrotron emission
\citep{Preece00,Kaneko06}
and suggesting an opaque source.

Accepting that the MeV peak of GRB spectrum forms inside an opaque jet
leads to so-called ``photospheric'' emission models.
The burst radiation is released 
where the jet becomes sufficiently transparent to scattering,
and its spectrum is mainly shaped by subphotospheric radiative processes.
Several versions of the photospheric model have been proposed over the years
(\citealt{Thompson94,Eichler00,MeszarosRees2000,GianniosSpruit07};
\citealt[][hereafter B10]{B10}; \citealt{Levinson12,ThompsonGill14}).
All share a key feature: the jet is {\it dissipative}, i.e. significantly heated as it 
propagates away from the central engine. 
This heating modifies the emitted photospheric 
radiation from simple blackbody emission. The resulting spectrum was shown to have a 
nonthermal shape that closely resembles the phenomenological Band function
(\citealt{Peer2006,Giannios2008}; B10; \citealt[][hereafter V11]{VBP11}).
It was proposed that the dissipative photosphere model provides a good
description to the observed spectra \citep{Ryde11}
and needs to be carefully tested against observations.

\subsection{Internal dissipation}

Four dissipation mechanisms have been proposed as sources of GRB emission: 
collisionless shocks \citep{ReesMeszaros94}, 
damping of Alfv\'en wave turbulence \citep{Thompson94}, 
magnetic reconnection \citep{DrenkhahnSpruit02},
and neutron collisions (B10).
Magnetic field and internal bulk motions provide the energy
reservoirs available for dissipation. 

The presence of strong internal motions in the jet is 
indicated by the observed 
variability of GRB radiation.
The central engine of the explosion is likely unsteady,
and additional variability is induced as the jet burrows its way through the progenitor
star and the cocoon produced by the jet-star interaction \citep{Lazzati09}.
This leads to multiple internal and recollimation shocks, which keep the jet hot and 
relatively slow when it emerges from the stellar progenitor. Thus, shock heating is 
expected to occur in
an extended range of radii and in an extended range of timescales, which is consistent 
with the observed broad power spectrum of variability
\citep{Beloborodov00,Morsony10}.

Additional evidence for dissipation at small radii is provided by the 
observed {\it photon number} emitted in GRBs.
In many GRBs, the central engine is unable to provide the observed photons, 
so additional photons must be produced in the expanding jet. 
Photon production is a direct consequence of dissipation at large optical depths 
(B13; \citealt{Vurm13}, hereafter V13;
see also \citealt{Eichler00,Thompson07}).
Observations also require that dissipation continues at least to the photospheric radius,
so that the released spectrum has a nonthermal shape. 
Therefore, in this paper we consider outflows which remain dissipative across 
a broad range of distances from the central engine, starting from the region inside 
the progenitor and extending to the jet photosphere and beyond. 

As long as baryons dominate the plasma inertia, dissipation of internal motions may be 
expected to heat the ions (and neutrons) in the first place. Efficient dissipation should 
give a typical energy of $\sim 1$~GeV per nucleon (its rest mass) in a relativistic jet.
Baryons themselves do not emit significant radiation, because of their large mass-to-charge 
ratio, however their energy can be passed to the electrons and radiated in the following ways:

\begin{enumerate}

\item 
Coulomb collisions gradually
pass energy from the hot ions to the thermalized 
electron/positron population (which is kept much colder by efficient radiative cooling).

\item
Inelastic (pion-producing)
nuclear collisions generate a non-thermal $\epm$ population with Lorentz factor 
$\ginj\sim m_\pi/m_e\sim 300$.

\item
Plasma motion through a radiation-mediated shock has a steep 
velocity gradient, leading to bulk Comptonization of photons by the electrons.
Bulk Comptonization could also result from plasma turbulence on small scales.

\item
Collective plasma processes in a collisionless shock suddenly heat the electrons to 
an ultra-relativistic temperature.

\end{enumerate}
All these processes are important in sub-photospheric internal shocks \citep{Beloborodov16}.
Electrons can also be directly heated by magnetic reconnection, 
tapping into the magnetic energy carried by the jet.

\subsection{Evolution of radiation in the expanding jet}

The energized electrons rapidly lose their energy to radiation via inverse Compton (IC)
scattering, synchrotron emission, and (at extremely high optical depths) through 
double Compton scattering and bremsstrahlung. The produced photons are redistributed 
in energy by Compton scattering and form the spectrum that eventually escapes at 
the Thomson photosphere $\Rph$ 
where the scattering optical depth $\tauT$ drops below unity.

Three relevant regions in the jet were described in B13:
\begin{enumerate}

\item The Planck zone ($r\lesssim 10^{10}$ cm, $\tauT \gtrsim 10^5$):
the density of the jet is sufficiently high to maintain blackbody radiation in 
detailed equilibrium with the thermalized plasma.

\item The Wien zone ($\tauT \gtrsim 10^2$): the dissipated heat is thermalized
into a Bose-Einstein photon distribution with a {\it finite chemical potential}.
The number of photons accumulated in the  
Wien peak attains its final value near the Wien radius,
beyond which Comptonization is unable to bring 
new generated photons to the spectral peak.

\item Unsaturated Comptonization zone ($\tauT \lesssim 10^2$): 
heating maintains a Compton parameter $y\sim 1$. The final non-thermal shape of the 
spectrum is produced in this region.

\end{enumerate}
The GRB radiation emerging at the photosphere
carries information about the entire expansion history of the jet.
Thus, the observed spectrum may be used to reconstruct dissipative processes 
hidden in the opaque region behind the photosphere.

It is best to view the formation of the observed spectrum as a problem of radiative 
transfer in an ultra-relativistic outflow \citep[][hereafter B11]{B11}.
The problem must be solved consistently with the flow dynamics and heating.
Boundary conditions should
be set sufficiently close to the central engine where 
the hot and dense plasma is capable of producing and thermalizing photons,
so that these processes can be explicitly followed by the transfer simulation.
Radiative transfer from the inner region
to and beyond the photosphere 
determines the radiation spectrum released by the jet.

\subsection{This paper}

Our main goal in this paper is the development of a global radiative transfer 
model for heated jets and the application of the model to observed GRBs.
The main novel feature of our simulations is the explicit inclusion of photon 
production reactions in the transfer problem. This requires one to start the simulation 
at a small radius, extremely deep below the photosphere. 

This paper focuses on jets whose energy is dominated by matter and radiation,
and a modest magnetization parameter $\epsB<1$ is assumed. We defer our study of 
magnetically-dominated jets to a future work; a recent discussion of radiative 
processes in magnetically dominated jets is found in 
\citet{GillThompson14,ThompsonGill14} and \citet{BeguePeer15}.

To keep the number of parameters to a minimum we employ a simple model of 
continuous internal dissipation
approximating the average heating rates (thermal and non-thermal) 
as power laws of radius. 
This may be a crude approximation to jet heating by multiple internal 
shocks or reconnection, 
however it allows us to study the global picture of the evolution of radiative 
processes with radius. As will be demonstrated below, the formation of
photospheric radiation
extends over several decades in radius, and
the transfer simulation allows us to study all stages of this process.
The small number of parameters in the model makes it useful for fitting the data.

The simulations presented in this paper are performed with an improved version 
of the kinetic code developed by \citet{VP09} and \citet{VBP11}.
It follows in detail all relevant radiative processes
expected in a heated jet, thermal and nonthermal.
The thermal processes include Comptonization of photons by the thermal plasma,
induced Compton scattering at low frequencies, cyclotron emission/absorption, 
bremsstrahlung, and double Compton scattering. The rates of all these processes
are well defined and accurately calculated for a given density, temperature, and 
magnetic field.

The injection of relativistic leptons by nuclear collisions
or by collisionless heating (Beloborodov 2016)
leads to synchrotron emission inside the jet, which is an important source of photons. 
Relativistic leptons also generate an IC $e^\pm$ cascade;
it is calculated in detail using exact cross sections. The density of $e^\pm$ pairs 
is governed by the rates of their creation and annihilation; accurate 
calculation of $e^\pm$ density is essential as it can strongly dominate over 
the electron-ion plasma density.

The paper is organized as follows.
Section~\ref{sec:setup}
describes the setup of the transfer 
problem and the physical processes involved. Sample models are presented 
in Section~\ref{sec:models}
where we explore how
the emitted spectrum is formed as the jet expands over several decades in radius,
how the emerging radiation is influenced by
the jet magnetization, acceleration history etc.
In Section~\ref{sec:fits},
we apply the model to three bright bursts
with high-quality spectral data: GRB~990123, GRB~090902B and GRB~130427A.
The results are discussed in Section~\ref{sec:disc}.

%###############################################################

\section{Setup of the transfer simulation}

\label{sec:setup}

\subsection{Dynamics of the jet}

We assume that the jet collimation occurs inside a characteristic radius
$\rcoll$ and its subsequent expansion  at $r>\rcoll$ may be approximated as conical.
In our models, $\rcoll$ is typically chosen around $10^{11}$~cm;
our simulations start 
at this radius. The jet is still accelerating at this stage, and its Lorentz factor 
saturates at a much larger radius; the acceleration is self-consistently calculated
as described below.

The jet variability is viewed as internal turbulent motions $\Delta \Gamma$ 
superimposed on a steady flow with Lorentz factor $\Gamma(r)$.
At any radius $r$, the internal motions can be divided into two parts: large-scale ``frozen''
fluctuations (length-scale larger than the local ``horizon'' --- the size of the casually 
connected region) and small-scale evolving fluctuations. This decomposition is similar 
to the analysis of primordial perturbations in cosmology. 
The small-scale fluctuations are dissipative, and they are 
assumed in our radiative transfer problem to be the
source of heat and nonthermal particles. 
The jet heating must be unsteady; however, we simplify the simulation by using
a smoothed, averaged heating rate. 
When calculating radiative transfer, we average out $\Delta\Gamma$ and 
view $\Gamma$ as a single-valued function of radius $r$ (or the 
comoving time of the steady background flow).
Then radiative transfer can be modeled in a steady-state approximation,
which significantly simplifies the calculations
(see B11 for discussion of this approximation).

The turbulent motions $\Delta\Gamma$ are 
only viewed as a reservoir of energy available for dissipation.
Then the total jet luminosity may be decomposed as
\begin{align}
L = \Lpl + \Lrad + \LB + \Lturb =const.
\label{eq:Lcomp}
\end{align}
Here $\Lpl(r)$ is the steady component of the plasma energy flow rate
separated from the turbulent component $\Lturb(r)$ (the reservoir);
$\Lrad(r)$ and $\LB(r)$ are the energy flows carried by radiation and magnetic field.
As the jet propagates, energy can be redistributed between the different components in 
Equation (\ref{eq:Lcomp}). The dissipated part of the turbulence reservoir 
gets converted to the plasma internal energy, 
which is immediately transferred to radiation via rapid cooling;
thus effectively there is a gradual transformation $\Lturb \rightarrow\Lrad$.
In the simulations presented below, we keep $\LB=\mbox{constant}$ for simplicity;
in reality, the magnetic component may have a 'reducible' component available for 
dissipation, e.g. via reconnection. 
However, when $\epsB=\LB/L\ll 1$ the dissipation of magnetic 
fields cannot serve as the main source of GRB radiation.

Radiation can efficiently accelerate the jet at early stages, when it dominates the 
jet energy. Thus we also expect the gradual transformation $\Lrad \rightarrow\Lpl$.
The dynamical equation for the jet Lorentz factor $\Gamma$ is given by 
(see B11),
\begin{align}
\frac{d\Gamma}{d r} = \sigmaT \Zpm \, \frac{4\pi I_1}{m_p c^3},
\label{eq:Gevol2}
\end{align}
where $4\pi I_1$ is the radiation flux measured in the rest-frame of the jet 
($I_1$ is the first moment of the local radiation intensity), and 
$Z_\pm$ is the number of electrons and positrons per proton.
Equation~(\ref{eq:Gevol2}) assumes $\Lpl\approx\Gamma\dotM c^2$;
this is a good approximation as long as
the ion temperature is non-relativistic. 

We use Equation~(\ref{eq:Gevol2}) in our simulations to calculate the self-consistent 
$\Gamma(r)$. At very large optical depths, where radiation is nearly isotropic and $I_1$
is small, an alternative form of the dynamical equation can be used,
which does not depend explicitly on $I_1$, see Equation~(\ref{eq:app:Gevol2_th}) 
in Appendix \ref{sec:app:dyn}.

\subsection{Kinetic equations}

At optical depths $\tauT\gtrsim 100$ the radiation field is nearly isotropic 
in the jet rest frame,
and the bulk of leptons are kept at a non-relativistic temperature.
The evolution of radiation under such conditions is well described by the Kompaneets 
equation. For a conical accelerating outflow it takes the form
\begin{align}
&\frac{1}{r^2\Gamma} \frac{\d}{\d\ln{r}} \left( r^2\Gamma \, \nocc \right) = 
\frac{r}{c\Gamma} (\jocc - c\knu \nocc) \nonumber \\
&+\frac{1}{x^2} \frac{\d}{\d x}
\left[ \tauT x^4 \left( \thetae\frac{\d \nocc}{\d x} + \nocc + \nocc^2 \right)
 + \frac{3-g}{3} \, x^3 \nocc
\right],
\label{eq:Komp}
\end{align}
where $\nocc$ is the photon occupation number, $x=h\nu/\me c^2$ is the dimensionless 
photon energy and $\thetae=\kB\Te/\me c^2$ is the 
dimensionless electron temperature.
The last term in square brackets accounts for adiabatic cooling of radiation, where
\begin{align}
    g = 1- \frac{d\ln\Gamma}{d\ln{r}}.
\end{align}
All quantities except $r$ and $\Gamma$ are measured in the instantaneous comoving frame of the jet.
The opacity $\knu$ and the source term $\jocc$ account for
all relevant radiative processes (see Section \ref{sec:radproc}) except thermal Comptonization,
which is  described by the Kompaneets equation itself.
Note that it is essential to include the induced $\nocc^2$ term in
the Kompaneets equation to accurately account for the number of synchrotron
photons that can be upscattered to the Wien peak (V13); neglecting this term 
would lead to an overestimation of 
the photon number in the Wien peak.

The deviation from isotropy develops at optical depths 
$\tauT\sim ~ \mbox{a few tens}$ (B11),
roughly near the Wien radius.
At larger radii one has to solve the full angle-dependent radiative transfer 
equation (\citealt{Mihalas80}; B11),\footnote{
    In contrast to the Kompaneets equation, \Eq~(\ref{eq:RTE})
    neglects induced downscattering, which 
    affects the rate of photon upscattering from low energies toward the Wien peak.
     This loss of accuracy is acceptable, because our simulation switches to \Eq~(\ref{eq:RTE})
    outside the Wien zone where upscattering has stopped
    populating the Wien peak with new photons.
    }
\begin{align}
&\frac{1}{r^2 \Gamma}\frac{\partial}{\partial\ln{r}} \left[ (1+\mu) r^2 \Gamma \Inu \right] = 
\frac{r}{\Gamma} (\jnu - \knu\Inu)	\nonumber \\
&+ (1+\mu)(1 - g\mu) \frac{\d\Inu}{\d\ln\nu}
- \frac{\d}{\d\mu} \left[ (1-\mu^2)(1+\mu) \, g\Inu
\right],
\label{eq:RTE}
\end{align}
where $\Inu$ is the comoving specific intensity, $\mu = \cos\theta$, 
$\theta$ is the (comoving frame) angle
relative to the radial direction, and $j_\nu$ is the emissivity.
This equation accurately describes the relativistic radiative transfer in 
three dimensions for spherically symmetric outflows 
with $\Gamma \gtrsim 10$.

The evolution of the electron/positron distribution is described by the kinetic equation
\begin{align}
&\frac{1}{r^2\Gamma} \frac{\d}{\d\ln{r}} \left[ r^2\Gamma \, \npm(p) \right] = 
\frac{r}{c\Gamma} (\jpm - c\kpm \npm(p)) \nonumber \\
& -\frac{\d}{\d p}
\left\{
\frac{r}{c\Gamma} \left[
\dotp\, \npm(p) - \frac{1}{2} \frac{\d}{\d \gamma}\left[ D \npm(p) \right]
\right]
- \frac{3-g}{3} \, p \,  \npm(p)
\right\},
\label{eq:elkin}
\end{align}
where $p=\sqrt{\gamma^2 - 1}$ is the electron/positron momentum in 
units of $m_ec$, and $\npm(p)$ is the electron/positron density per 
unit momentum interval.
The terms $\dotp$ and $D$ account for heating/cooling and diffusion in momentum space
due to radiative processes 
and Coulomb collisions.
The emission term $\jpm$ includes
the injection of new pairs due to non-thermal dissipation 
and photon-photon absorption $\gamma+\gamma\rightarrow e^++e^-$,
and $c\kappa_\pm n_\pm(p)$ is the pair annihilation rate.
All these terms are accurately calculated as described in
\citet{VP09} and \citet{VBP11}.
The last term in curly brackets describes adiabatic cooling.

In the simulation
the photon equation is solved together with the kinetic equations for electrons/positrons
and the dynamical equation (\ref{eq:Gevol2}) for $\Gamma(r)$.
The photon and lepton equations are coupled via the interaction terms 
$\jocc$ ($\jnu$), $\knu$, $\jpm$, $\kpm$, $\dotp$ and $D$.
The switch from the Kompaneets equation (\ref{eq:Komp}) to the radiative transfer 
equation (\ref{eq:RTE}) is made at $\tauT=100$; at this stage either of these equations 
is sufficiently accurate,
allowing for a smooth transition.
The evolution is followed to $\tauT\ll 1$.
The simulation is typically stopped at $\tauT \lesssim 0.1$ where
the obtained radiation field is transformed to the observer frame
and assumed to escape.

\subsection{Radiative processes and photon generation}

\label{sec:radproc}

Radiative processes in the expanding jet determine the observed GRB spectrum
and require accurate calculations at all radii, including very opaque regions far below 
the photosphere.
In particular, the position of the spectral
peak is controlled by photon generation below the Wien radius,
where photons are produced at low energies and up-scattered to the peak via saturated Comptonization.
The photon generation mechanisms can be categorized a follows (B13, V13):
\begin{enumerate}
\item Thermal: photons are generated by Maxwellian leptons. 
Thermal radiative processes operate efficiently only in the Planck zone
($\tauT\gtrsim 10^5$).
These processes include double Compton scattering, bremsstrahlung (less efficient than 
double Compton unless the jet is strongly clumped), and cyclotron emission.
The efficiency of cyclotron emission can be similar to that of double Compton scattering 
if the jet is strongly magnetized.
  \item Nonthermal: photons are generated by relativistic nonthermal electrons/positrons.
  The main radiative process is synchrotron emission.
  It continues to operate outside the Planck zone, and the number of produced 
  photons typically {\it increases} with radius, 
  as long as the non-thermal dissipation channel remains active.
\end{enumerate}

The rates of the thermal processes are given in B13 and V13. The implementation of 
nonthermal synchrotron emission is similar to that in V13, with one significant
extension: 
we are considering jets with moderate magnetization $\epsB \sim 10^{-3}-10^{-1}$
where pair-photon cascades produce a larger number of secondary nonthermal particles (see below).
In contrast,
V13 considered magnetizations close to equipartition where pair cascades are suppressed.

\subsection{Heating rate}

The total (thermal and non-thermal) heating rate satisfies
\begin{align}
\frac{d\Lh}{d\ln{r}} \equiv
\frac{d\Lth}{d\ln{r}} + \frac{d\Lnth}{d\ln{r}}
= -\frac{d\Lturb}{d\ln{r}}.
\label{eq:dissip}
\end{align}
The simulations presented below assume that the fractional luminosities dissipated 
via the thermal and nonthermal channels follow a power law,
\begin{align}
\label{eq:epsth}
   \epsth &\equiv  \frac{1}{L}\frac{d\Lth}{d\ln{r}} =
\ethinit \left( \frac{r}{\rcoll} \right)^{\kth},  \\
\label{eq:epsnth}
\epsnth &\equiv  \frac{1}{L}\frac{d\Lnth}{d\ln{r}} = \enthinit
\left( \frac{r}{\rcoll} \right)^{\knth}.
\end{align}
The corresponding heating and injection rates in Equation (\ref{eq:elkin}) are
\begin{align}
\dotp_{\rm h} = \frac{\gamma}{p}\dotg_{\rm h} = \frac{\gamma}{p}\frac{\epsth L}{4\pi \me c^2 r^3 \Gamma (n_- + n_+)},
\end{align}
and
\begin{align}
j_{\pm, {\rm inj}} = \frac{\epsnth L}{4\pi \me c^2 r^3 \Gamma} \, \frac{\delta(p - \pinj)}{2\ginj},
\end{align}
where $n_-$ and $n_+$ are the electron and positron densities, respectively,
$\pinj$ is the dimensionless momentum of the injected pairs,
and $\ginj=(1+\pinj^2)^{1/2}$ is their Lorentz factor.

\subsection{Parameters of the model}

The GRB models presented below have the following set of parameters:
\medskip

\noindent
(1) Total jet luminosity $L$ (isotropic equivalent). \\
\noindent
(2) Energy per baryon $\eta=L/\dot{M}c^2$. \\
\noindent
(3) Magnetization $\epsB=L_B/L$. \\
\noindent
(4) The jet Lorentz factor at the radius $\rcoll$ where the simulation starts; $\rcoll$ 
is chosen near the radius of the jet breakout 
from the progenitor star.\\
(5) The jet energy per photon, i.e. 
the initial $\Epk$.
In our models the jet is initially photon starved, i.e. we choose a high $\Epk\approx 10$~MeV.
This is done to demonstrate that even in weakly magnetized jets synchrotron emission
can efficiently produce photons. Then the predicted GRB spectrum weakly
depends on the choice of initial $\Epk$.
\\
\noindent
(6) The efficiencies of thermal and nonthermal heating, $\epsth$ and $\epsnth$.
In most of our models heating is assumed to have a flat distribution over $\ln r$ 
($\kth=\knth=0$ in Equations~(\ref{eq:epsth}) and (\ref{eq:epsnth})); 
we will also consider a few variations in the heating history of the jet
if this is required to fit the observed GRB spectra.

The non-thermal lepton injection energy
is fixed at $\ginj = 300 \approx m_\pi/m_e$, as expected for 
collisional heating; its exact value weakly affects the results.

%###################################################################

\section{Sample models}

\label{sec:models}
  
  \subsection{Overview of the spectral evolution}

\begin{figure}[t]
  \plotone{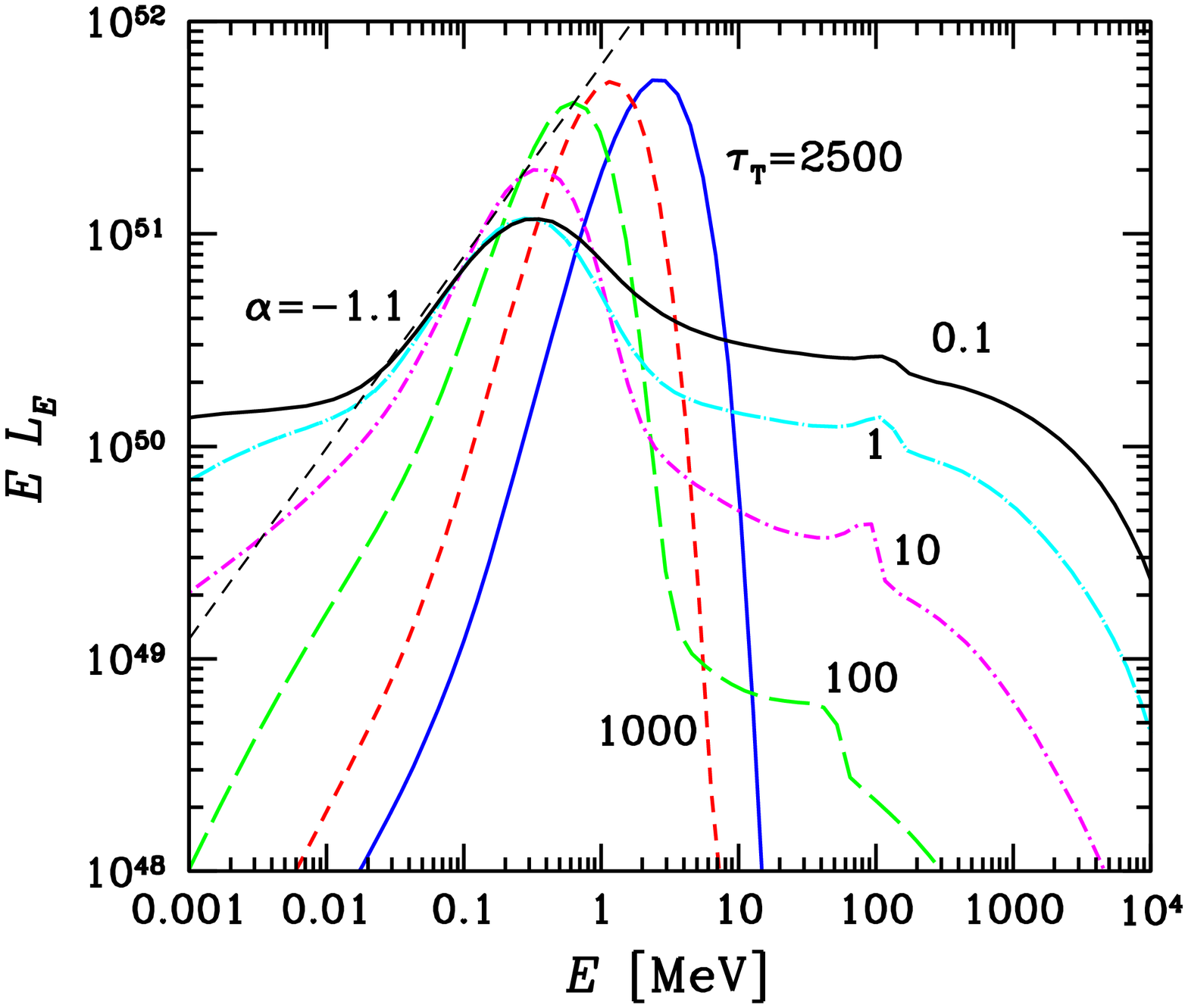}
  \caption{Evolution of the radiation spectrum  
 carried by the jet, from high to low optical depths $\tauT$.
 This sample model has the following parameters:
  $L=2\times 10^{52}$ erg/s, $\eta=190$,
  $\rcoll=10^{11}$~cm, $\gamma_0=300$, 
  $\epsth+\epsnth=0.05$, $\epsth/\epsnth=1$,
  $\Gamma(\rcoll) = 20$, magnetization
  $\epsB= 10^{-2}$.
  The spectra are measured in the rest frame of the central engine and
  not corrected for a cosmological redshift.}
  \label{Fig:evol}
  \end{figure}

Figure \ref{Fig:evol} illustrates the evolution of the 
photon spectrum within 
a dissipative jet as the radiation is carried from highly opaque regions to transparency.
There are two main stages of the spectral evolution:

(1) Generation of photons below the Wien radius ($\tauT\gtrsim 10^2$)
and their Comptonization to the Wien peak. 
The growth of photon number 
in the Wien peak results in its shift to lower energies.
The high-energy tail is suppressed at $r<\rW$ for two reasons.
First, the electrons are kept approximately at the Compton temperature,
i.e. $3\kB\Te \approx \Epk/\Gamma$.
Thus thermal Comptonization is 
hardly capable of populating the photon spectrum above $\Epk$,
leading to an exponential cutoff at $\Epk$. 
Secondly, the nonthermal radiation from IC
scattering by relativistic pairs
does not survive at $\tauT\gg 1$
--- it is reprocessed to lower energies via the pair-photon
cascade and down-scattering by the thermal pairs.

(2) Broadening of the spectrum by unsaturated Comptonization
outside the Wien radius, leading to a nonthermal shape of the spectrum.
The broadening first affects the low-energy slope $\alpha$, which
begins to soften near the Wien radius and attains its final value near the photosphere.
The resulting photon index $\alpha\sim -1$ depends on the heating history 
at $r>\rW$, in particular on the nonthermal dissipation channel, as
the unsaturated Comptonization of the fresh low-energy synchrotron photons 
governs the softening of $\alpha$.
Without nonthermal dissipation the low-energy spectral slope would be
much harder.  

The photospheric spectrum also broadens at high energies.
The high-energy tail is mostly built around and above the photosphere,
due to two effects, thermal and nonthermal.
Outside the Wien radius the electron temperature rises 
significantly above $\Epk/\Gamma$ (see below) and thermal Comptonization 
begins to produce photons with energies $E>\Epk$.
As the optical depth drops, the nonthermal high-energy 
component becomes increasingly prominent, especially
in weakly magnetized jets. 
Then the overlapping thermal and nonthermal Comptonization
components together form an extended high-energy spectrum
which can easily be mistaken for a single emission component
(see also B10; V11).
In contrast, in strongly magnetized jets with $\epsB\sim 1$, the pair cascade and
nonthermal Comptonization are reduced, leading to a different spectral shape at high
energies (see Figure~5 in V11).

%%%%%%%%%%%%%%%%%%%%%%%%%%%%%%%%%%%%  
%%%%%%%%%%%%%%%%%%%%%%%%%%%%%%%%%%%%
     
It is worth noting that the overall
spectrum has a distinctly nonthermal shape already
well below the photosphere (at $\tauT\gtrsim 10$).
In particular, the low-energy slope is significantly softer than
the Planck (or Wien) spectrum.  
Even if dissipation stopped completely at 
$\tauT\sim 10$, a thermal-looking prompt emission would not be expected;
instead, the emerging spectrum would resemble a cutoff power law.

A thermal-looking GRB spectrum would only be produced
in a special case when all dissipation is confined to very high optical depths.
Even in this case, however, the transfer effects would soften $\alpha$
from the thermal slope to $\alpha\sim 0.4$
(B10; see also \citealt{PeerRyde11}).

\subsection{$e^\pm$ cascade and pair loading}

\label{sec:casc}

When the jet magnetization is weak, $\epsB<0.1$,
the relativistic $e^\pm$ pairs injected by the nonthermal dissipation channel
lose their energy primarily
through IC scattering of photons from the quasi-thermal (Wien) peak.
The upscattered photons with energies above $\me c^2$ in the jet frame immediately
convert to secondary $\epm$ pairs through photon-photon collisions.
The created particles also upscatter photons etc.,
leading to a pair-photon cascade that populates the jet with
secondary $\epm$ pairs which can dominate the plasma.

The distribution of non-thermal leptons resulting from the cascade can be expressed as
(cf.~Equations~(18) and (19) in V11)
\begin{align}
  \Ne
  (\gamma) = \frac{3Y\gcascmin^{\alphacasc}}{4c\sigmat} \frac{\epsinj}{(\epsB + \epsrad)}     
  \frac{c\Gamma}{r} \
  \gamma^{-(\alphacasc+2)},
\label{eq:Ne:casc}
\end{align}
where
$\epsB = L^{-1} d\LB/d\ln{r}$, $\epsrad=L^{-1} d\Lrad/d\ln{r}$,
and $Y\sim 0.1$ is the pair yield
\citep{Sve87}.
The saturated cascade turns off at $\gcascmin\approx (\me c^2\Gamma/\Epk)^{1/2}\sim 10$;
electrons at $\gamma<\gcascmin$ are unable to upscatter photons from the Wien peak to MeV energies for further pair generation.
The pair yield of the cascade generated by a 
primary particle with Lorentz factor $\ginj$ may be expressed as $Y=\Ms/\ginj$, where 
$\Ms$ is the pair multiplicity, i.e. the number of secondary $e^\pm$ per primary particle.
The power-law index $\alphacasc\approx\ln{\Ms}/(\ln{\ginj} - \ln{\gcascmin})$ characterizes the 
additional steepening effect of the $e^\pm$ cascade on   
the standard cooling distribution $\Ne(\gamma)\propto \gamma^{-2}$.

\begin{figure}[t]
  \plotone{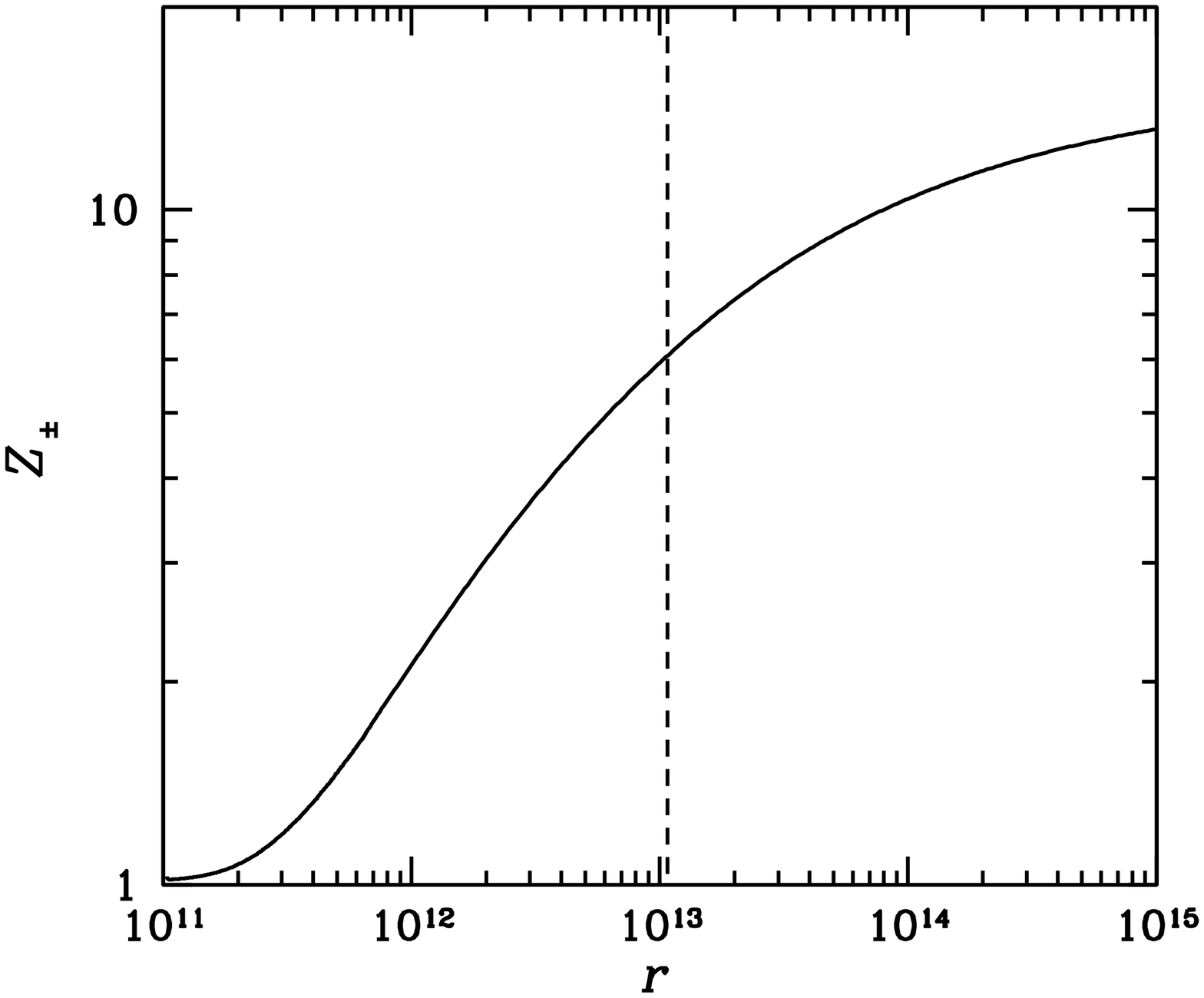}
  \caption{Evolution of pair-loading with radius
  in the simulation shown in Figure~\ref{Fig:evol}. 
  The vertical dashed line indicates the freeze-out radius where $\taupm=16/3$.
  }
  \label{Fig:pload}
  \end{figure}

Below the photosphere the pairs are in creation-annihilation equilibrium.
Their density can be expressed as 
\begin{equation}
  \Ne=\Zpm \nprot,
\end{equation}
where the proton density is
\begin{align}
\nprot = \frac{\dotM}{4\pi\mprot c \, r^2\Gamma},
\end{align}
and the pair loading factor $Z_\pm$ is given by 
(Appendix~\ref{sec:app:pload}; see also B10; \citealt{ThompsonGill14})
\begin{align}
\Zpm = \left(\frac{16 \, Y\epsinj L}{3\, \taup \, \Gamma\dotM c^2} \frac{\mprot}{\me} +1 \right)^{1/2}
\propto (r\Gamma)^{1/2}.
\end{align}
The last proportionality is valid if $\Zpm\gg 1$, i.e. if pairs
dominate over the electron-ion plasma. This condition is met when the Thomson 
optical depth associated with $n_p$ 
is sufficiently low,
\begin{align}
\taup 
  =\frac{ \sigmaT \nprot \, r}{\Gamma} 
  \ll \frac{16}{3}\frac{\mprot}{\me}\frac{Y\epsinj L}{\Gamma\dotM c^2}.
\end{align}
This condition may not be satisfied in a region deep below the photosphere.
The expanding jet becomes increasingly pair dominated in the subphotospheric region, provided that 
the nonthermal dissipation channel is active (Figure \ref{Fig:pload}).

%%%%%%%%%%%%%%%%%%%%%%%%%
%%%%%%%%%%%%%%%%%%%%%%%%%

Pair annihilation freezes out
when the annihilation time, $t_{\rm ann} = 16/(3 c\sigmaT\npm)$,
exceeds the expansion time $t_{\rm dyn}=r/(c\Gamma)$, 
which occurs at $\taupm = 16/3$. The pair-loading 
factor at this time is
\begin{align}
\Zpm = \frac{\mprot}{\me} \frac{\epsinj Y L}{\Gamma\dotM c^2},
\end{align}
where $\Gamma$, $\epsinj$ and $Y$ are taken at the freezeout radius. Typically, 
we find $Z_\pm \sim 10$
for moderate nonthermal injection rate $\epsinj\sim 0.1$.
As shown in Appendix~\ref{sec:app:pload} and seen in Figure \ref{Fig:pload}, 
pair creation beyond the freezeout radius can increase 
$\Zpm$ by a modest factor (at most logarithmically in $r$).

The pair loading increases the photospheric radius by 
the factor of $\Zpm$ compared to the pair-free jet.
Pair loading is significantly reduced if the jet magnetization $\epsB$ is above a few percent,
since synchrotron cooling 
becomes competitive with Compton cooling, inhibiting
 the pair-photon cascade and lowering $Y$ (V11).

\subsection{
Building up the photon number in the Wien peak via synchrotron emission}

A fraction of the non-thermal lepton energy is 
converted to low-energy synchrotron radiation.
If the magnetization is weak, this fraction is small, as
the synchrotron losses of high-energy particles
are small compared to their IC losses.
While the energy budget of synchrotron radiation is small,
the {\it number} of generated synchrotron photons 
is substantial, because of their low energies.
Furthermore, these soft photons can gain energy 
from the thermal plasma through Comptonization. 
Below the Wien radius $\rW$ many of them are Comptonized to the Wien peak and 
eventually dominate the peak, significantly increasing its photon number and shifting
$\Epk$ to lower energies. This shift is required by energy conservation (energy is shared between
more photons) and accomplished through Compton cooling of the thermal plasma
by the synchrotron photons.

The number of synchrotron photons that can be upscattered to the 
Wien peak depends on the competition 
of Compton upscattering against synchrotron reabsorption
and induced Compton downscattering (V13).
This competition defines a critical photon energy $E_0$
(and a corresponding $\epm$ Lorentz factor $\gamma_0$),
above which upscattering dominates over other processes.
The calculation of $\gamma_0$, $E_0$, and the number of synchrotron photons
emitted above $E_0$ is given in Appendix~\ref{sec:app:phprod}.

At $r<\rW$ all synchrotron photons emitted at $E>E_0$ end up in the Wien peak.
Most of them are emitted near $E_0$ by leptons with Lorentz 
factors $\gamma\sim\gamma_0$
(the emissivity scales as $E^{(-p+1)/2}$, where $p\ge 2$). 
Therefore, the rate at which photons are accumulated in the peak 
[cm$^{-3}$ s$^{-1}$]
is roughly proportional to the number of leptons with $\gamma\sim\gamma_0$,
\begin{align}
    \dotNsynch \approx \frac{\sigmaT\me c^2 B}{3eh} \, \gamma_0
    \Ne(\gamma_0).
\label{eq:dotNph1}
\end{align}
The number of photons accumulated in the Wien
peak is finalized near $\rW$. 
The resulting flow of photon number through the sphere $4\pi \rW^2$ (isotropic
equivalent) is given by
\begin{align}
\label{eq:int}
   \dotNph
    = \int^{\rW} 4\pi r^3 \, \dotNsynch(r) \, d\ln{r}.
\end{align}
The photon number carried in the spectral peak is weakly changed outside the 
Wien radius; this number is eventually released at the photosphere.

The secondary $\epm$ pairs affect photon production by increasing both 
$\Ne$ and $\gamma_0$;
the net effect is positive, i.e. the pair cascade enhances photon generation.
Pair loading also increases $\rW$ by increasing the flow opacity;
this allows a longer time for the accumulation of photons in the 
Wien peak.

In contrast to thermal bremsstrahlung or double Compton effect, the 
synchrotron photon production increases with radius and the
integral in \Eq~(\ref{eq:int}) peaks near $\rW$. This behavior was seen for pair-free outflows
in V13 and remains true in the presence of pair cascades
(see Appendix~\ref{sec:app:phprod} for analytical estimates). Our numerical simulations confirm
that a substantial fraction of the photons accumulated in
the spectral peak originates near the outer boundary of the Wien zone.
This can be seen in Figure \ref{Fig:evol}:
the decrease of $\Epk$ due to continuing photon supply
to the peak ends around
$\tauT\lesssim 100$, which approximately corresponds to $\rW$.

As long as the nonthermal dissipation channel remains active,
the production of synchrotron photons continues 
beyond the Wien radius. This results in the
soft ``excess'' seen below a few$\times 10$~keV in 
the emitted spectrum (Figure~\ref{Fig:evol}, see also V11).
More importantly, unsaturated Comptonization of these photons plays 
a key role in determining the spectral slope below the peak,
which is discussed below.

\subsection{Comptonization}
\label{sec:Compton}

The saturated Comptonization
at $r<\rW$ maintains a quasi-equilibrium between radiation and thermal plasma.
The photon spectrum around the peak has a Wien shape, 
with small excesses at lower and higher energies
due to synchrotron and non-thermal IC
emission, respectively.
At $r>\rW$, the thermal Comptonization gradually switches to the unsaturated regime,
where the Compton $y$-parameter remains close to unity, as long as the heating operates.

The $y$-parameter may be evaluated by considering the evolution of the radiation luminosity
at $\tauT\gg 1$ (see Appendix \ref{sec:app:dyn}),
\begin{align}
\frac{d\Lrad}{d\ln{r}} = - \frac{2g}{3}\Lrad + \frac{4}{3} \frac{d\Lh}{d\ln{r}},
\label{eq:Lrad1}
\end{align}
where the first and second terms on the right hand side account for adiabatic cooling and dissipation, respectively.
Assuming $g = 1- d\ln\Gamma/d\ln{r} \approx \mbox{const}$ and the scaling
\begin{align}
\frac{d\Lh}{d\ln{r}} \propto r^{\kh},
\end{align}
one can solve Equation (\ref{eq:Lrad1}) for $\Lrad$,
\begin{align}
\Lrad = \frac{4}{3\kh + 2g} \, \frac{d\Lh}{d\ln{r}}.
\label{eq:Lrad}
\end{align}
The (thermal) heating of electrons is balanced by Compton losses,
\begin{align}
\frac{d\Lh}{d\ln{r}} =
\frac{3}{4} (y - \yC) \Lrad,
\label{eq:LhCompt}
\end{align}
where $y = 4\tauT\thetae$ and $\yC = 4\tauT\thetaC$ are the Compton parameters corresponding to the electron and Compton temperatures, respectively.
Substituting the solution (\ref{eq:Lrad}) into the heating-cooling balance 
(\ref{eq:LhCompt}), one finds that
the Compton parameter relaxes to
\begin{align}
y - \yC = \kh + \frac{2g}{3}.
\end{align}
Thus $y-\yC$ remains close to unity
if the heating rate does not decline
significantly faster than $d\Lh/d\ln{r}\propto r^0$.
At the Wien radius $y\approx \yC \approx \mbox{a few}$ and $\thetae\approx\thetaC$.
At larger radii
the electron temperature $\thetae$ increases above the Compton temperature $\thetaC$, 
as Comptonization gradually switches to the unsaturated regime 
with $y \approx \kh + 2g/3$.

The region with $y\sim 1$ can extend over several 
expansion times. Therefore, the effective net
Compton parameter can be substantially larger than unity
and result in significant redistribution of photons,
in particular above the spectral peak. 
The nonthermal high-energy tail is built in this regime,
with an energy content comparable to that of the radiation near the peak.
In addition to thermal Comptonization, 
the pair photon cascade initiated by the injected non-thermal 
leptons can extend the non-thermal tail to the GeV range.

The final low-energy slope of the photospheric spectrum is mainly formed at 
radii $r\simgt\rW$, in the intermediate regime with $y\sim~\mbox{a few}$.
It is shaped by the combination of photon Comptonization from low energies toward
the peak, diffusion in energy space, and adiabatic cooling.

Some insight into the development of the low-energy slope 
is provided by the quasi-steady 
solution of the Comptonization problem at a given radius 
in the spectral range $E_0 < h\nu < \kB\Te$.
The solution is found by setting the second term on the 
right side of Equation~(\ref{eq:Komp}) to zero.
Neglecting induced scattering, one finds\footnote{The actual spectrum at a given $r$
          has time to relax to the steady state solution only if $y\gg 1$.
          When $y\sim\mbox{a few}$, Equation~(\ref{eq:Compt:ss}) only 
          shows the qualitative behavior of the spectrum, but does not give
          its exact shape.
}
\begin{align}
   \Inu\propto \mbox{C} + \nu^{a}e^{-h\nu/\kB\Te}, 
\label{eq:Compt:ss}
\end{align}
where $C=~\mbox{const}$ and $a = 3-4(3-g)/(3y)$.	 
In the strongly saturated regime $y\gg 1$, 
the last term in \Eq~(\ref{eq:Compt:ss}) gives the Wien spectrum with $a\approx3$.
The other term $\Inu \propto \nu^0$ 
results from the steady photon flux in the energy space toward the Wien peak.

The relative amplitudes of the two components in $\Inu$ depend
on the generation rate of low-energy photons and 
the number of photons already accumulated in the Wien peak. 
As $y$ decreases, the Wien peak becomes  
less pronounced relative to the $\Inu\propto \nu^0$ component.
The spectral hardening toward the peak becomes weaker,
because $a$ decreases from its saturated value.
In addition, the peak itself is broadened by the competition
between Comptonization and adiabatic cooling.

Our detailed numerical simulations confirm
that the average spectrum approaches $\Inu \propto \nu^0$ 
below the peak for a broad range of parameters.
The most important requirement for this behavior
is the existence of a low-energy photon source between $\rW$ and $\rph$.
The spectrum $\Inu \propto \nu^0$ corresponds to the photon index $\alpha=-1$, 
which coincides with the average photon index observed in GRBs
\citep{Kaneko06}.

\subsection{The role of the early acceleration stage and magnetization}

\begin{figure}[t]
  \plotone{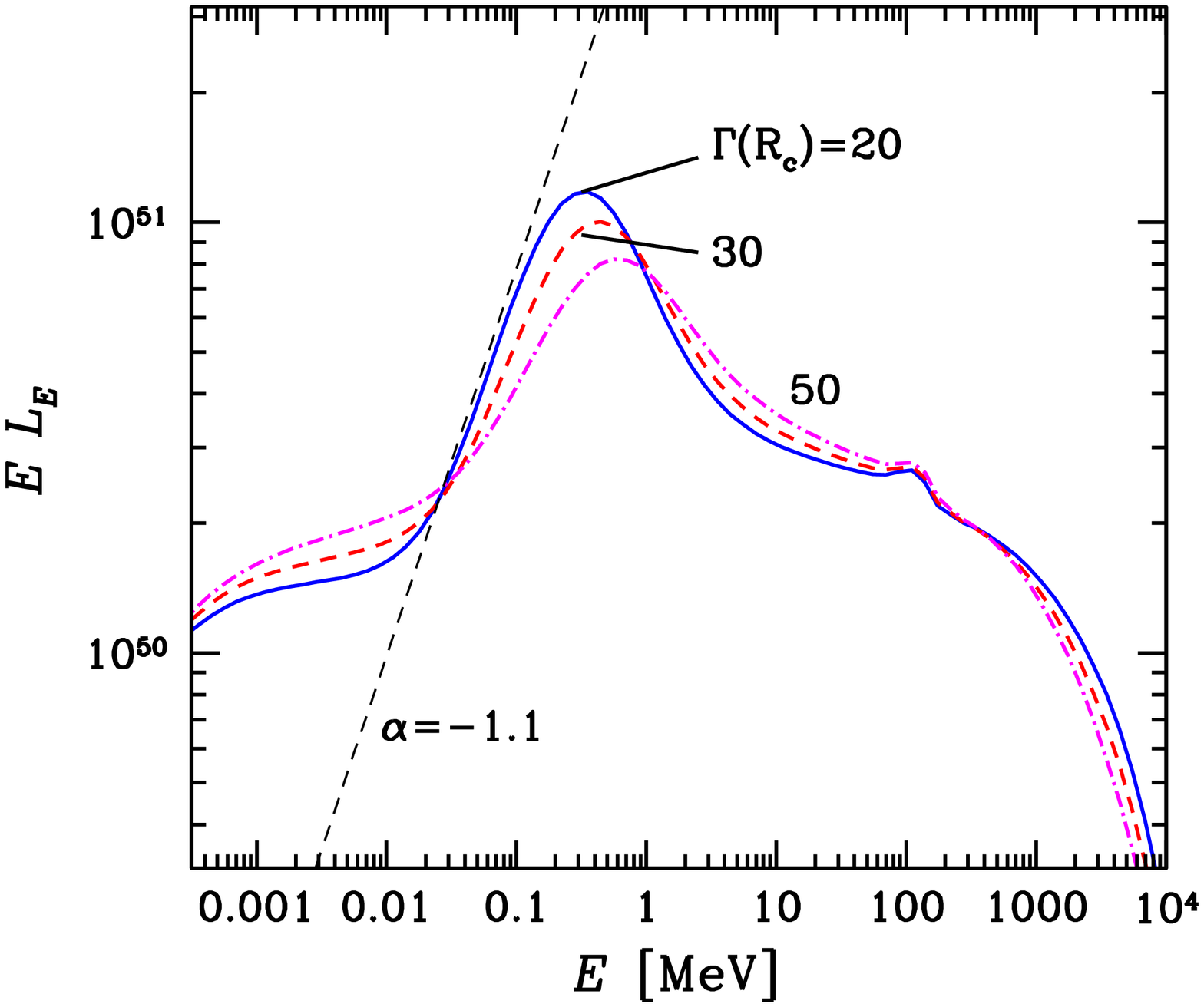}
  \caption{Effect of
   different Lorentz factors at $\rcoll$ 
   on the emitted spectrum. 
   Other parameters are the same as in Figure \ref{Fig:evol}.
   }
  \label{fig:spectra:Gamma}
  \bigskip
  \end{figure}

Figure \ref{fig:spectra:Gamma} shows the response of the final 
(observed) spectrum to variations in the Lorentz factor at the collimation radius $\Gamma(\rcoll)$.
Lowering $\Gamma$ at $r<\rW$
implies a higher density and increases the efficiency of photon production.
In addition, $\rW$ is increased, which increases the number of synchrotron photons
reaching the spectral peak (see Equation \ref{eq:app:Nsynch}).
Both effects lead to a lower $\Epk$.
  
   %%%%%%%%%%%%%%%%%%%%%%% 
 %%%%%%%%%%%%%%%%%%%%%%% 

The spectra become more narrow in jets that are still significantly accelerating
between $\rW$ and $\rph$, i.e. those with a larger disparity between $\Gamma(\rcoll)$
and the final $\Gamma$.
During the accelerating stage radiation carries a large fraction of the 
jet energy, which makes it less susceptible to spectral
redistribution/broadening by dissipation.
In addition, the ratio $\rph/\rW$ is smaller if the jet is still accelerating 
outside $\rW$, leaving less time for broadening the spectrum.
  
%%%%%%%%%%%%%%%%%%%%%%%%%%%%%%
    \begin{figure}[t]
  \plotone{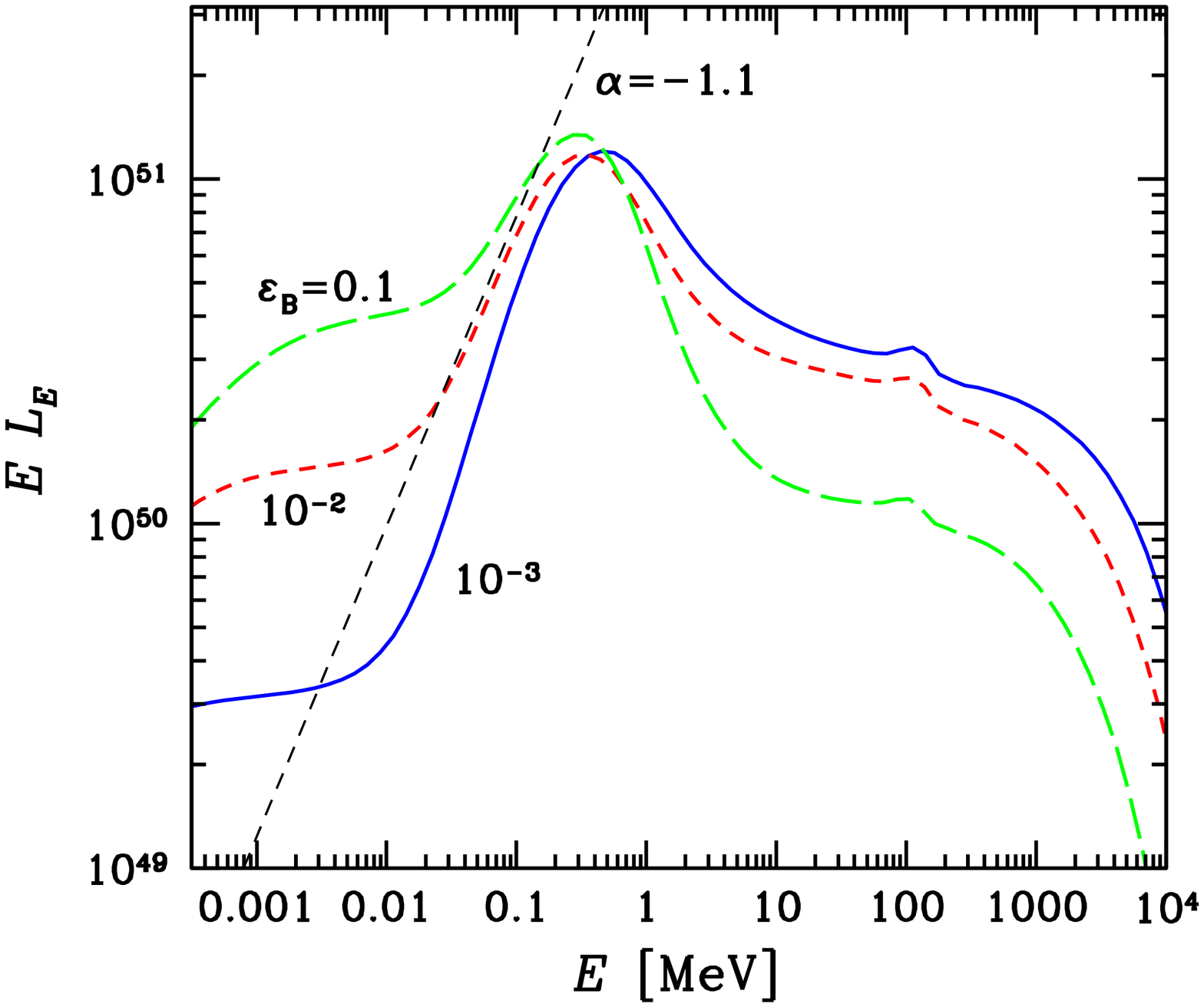}
  \caption{Effect of different magnetizations. Other parameters are the same as in Figure~\ref{Fig:evol}.}
  \label{fig:spectra:eB}
  \end{figure}
%%%%%%%%%%%%%%%%%%%%%%%%%%%%%%

Figure \ref{fig:spectra:eB} shows the spectra for different jet magnetizations.
With increasing $\epsB$ the peak position shifts to lower energies,
as more synchrotron photons are produced.
This effect is partially offset by the suppression of pair loading 
(see Section~\ref{sec:casc} and V11), which reduces the Wien radius 
and the number of synchrotron-emitting particles.
  
The spectral shape exhibits a characteristic behavior as the magnetization is increased:
more low-energy synchrotron photons tend to make the spectrum softer below the peak;
a stronger low-energy ``excess'' is also produced.
Above the peak the slope becomes steeper as synchrotron 
losses reduce IC emission from high-energy particles.
Synchrotron losses also inhibit pair cascade, which weakens and hardens the high-energy
spectrum. As a result, the nonthermal IC emission
creates a stronger upward curvature in the spectrum around 10~MeV.

%##################################################################  
  
\section{Fits to prompt GRB data}

\label{sec:fits}

To test our transfer simulations against observations, we chose three bright GRBs 
with good 
spectral data and distinct spectral shapes: GRB~990123, GRB~090902B, and GRB~130427A.
Good spectral fits were previously obtained with phenomenological models 
combining Band function \citep{Band09}, a power law, and sometimes 
a Planck component. Our model will pass the test if it is able to reproduce the fits.
Formal fitting utilizing an appropriate goodness of fit statistic as well as instrument response matrices is deferred to a future work.

\begin{figure}[t]
\plotone{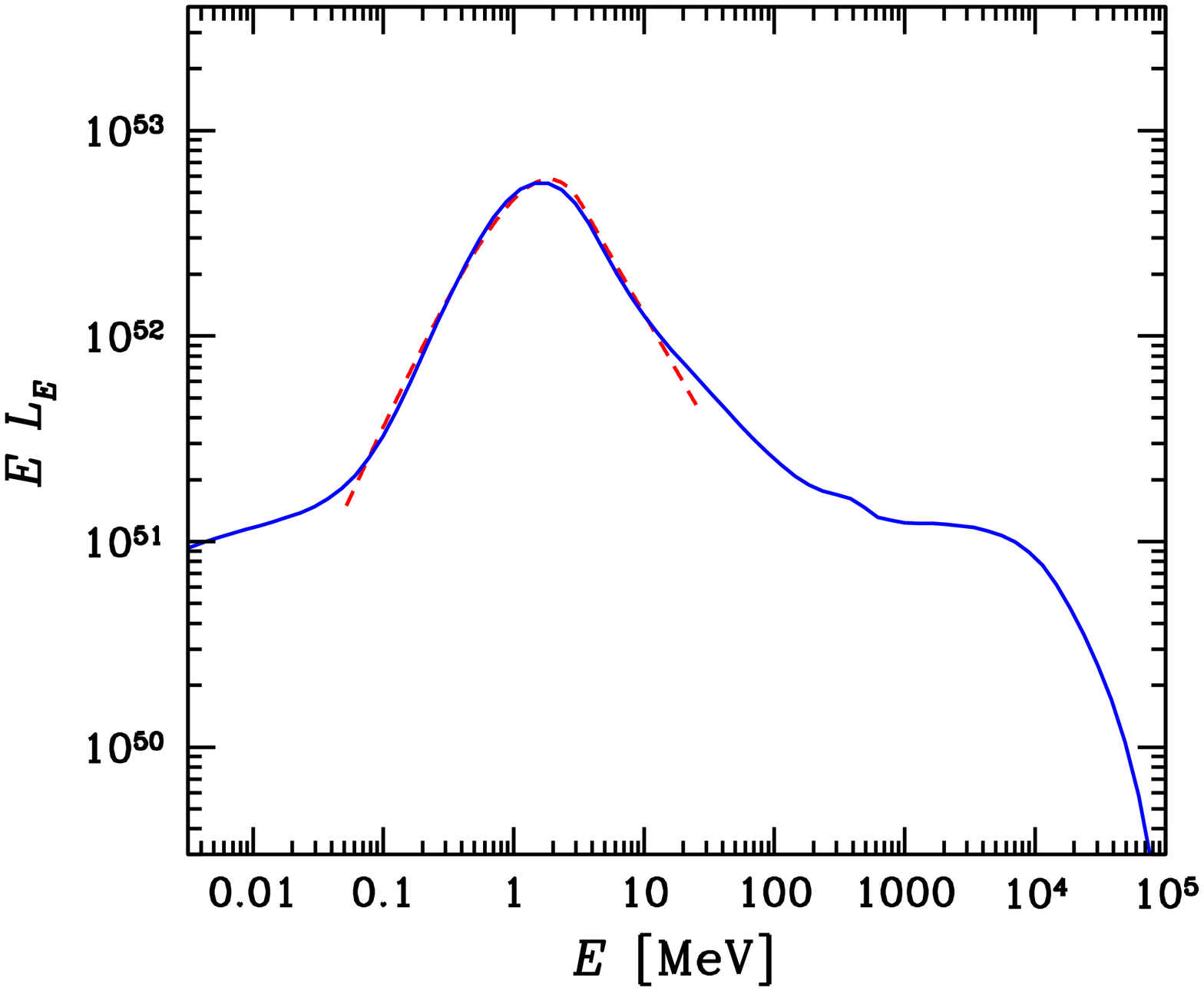}
\caption{Simulated spectrum of GRB 990123 (solid line), and a Band fit (dashed line, \citealt{Briggs99}). Jet parameters:
Magnetization $\epsB=0.018$,
simulations starting (collimation) radius $\rcoll=3\times 10^{10}$~cm, initial Lorentz factor $\Gamma(\rcoll)=35$,
initial number of photons per baryon $n_{\rm ph}(\rcoll)/n_{\rm p}(\rcoll) = 5\times 10^4$,
terminal Lorentz factor $\Gamma_{\rm f} = 500$.
The heating parameters are
$\ethinit+\enthinit=0.26$, $\ethinit/\enthinit=4.7$,
$\kth=\knth=-0.19$.
The heating proceeds until $\tauT=0.01$.
}
\label{Fig:990123}
\end{figure}

\begin{figure}[t]
\plotone{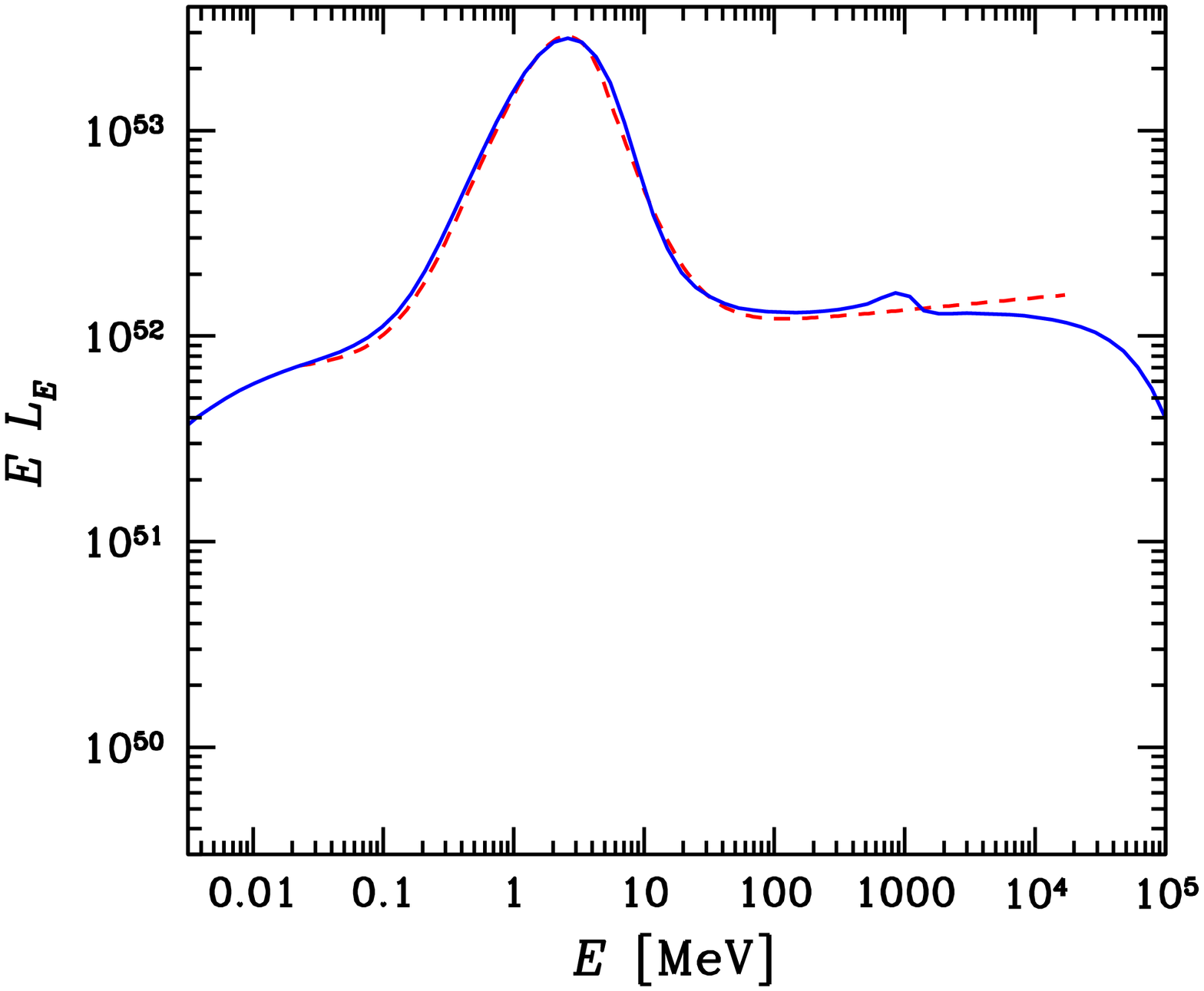}
\caption{Simulated spectrum of GRB 090902B (solid line), and a Band$+$power-law fit (dashed line, \citealt{Abdo_090902B_l}, bin b).
Jet parameters:
$\epsB=0.012$,
$\rcoll=3\times 10^{10}$~cm, $\Gamma(\rcoll)=70$, $n_{\rm ph}(\rcoll)/n_{\rm p}(\rcoll) = 10^5$,
$\Gamma_{\rm f} = 1220$;
Heating parameters
$\ethinit+\enthinit=0.12$,
$\ethinit/\enthinit=0.85$, $\kth=\knth=-0.25$.
The heating proceeds until $\tauT=0.03$.
}
\label{Fig:090902B}
\end{figure}

\begin{figure}[t]
\plotone{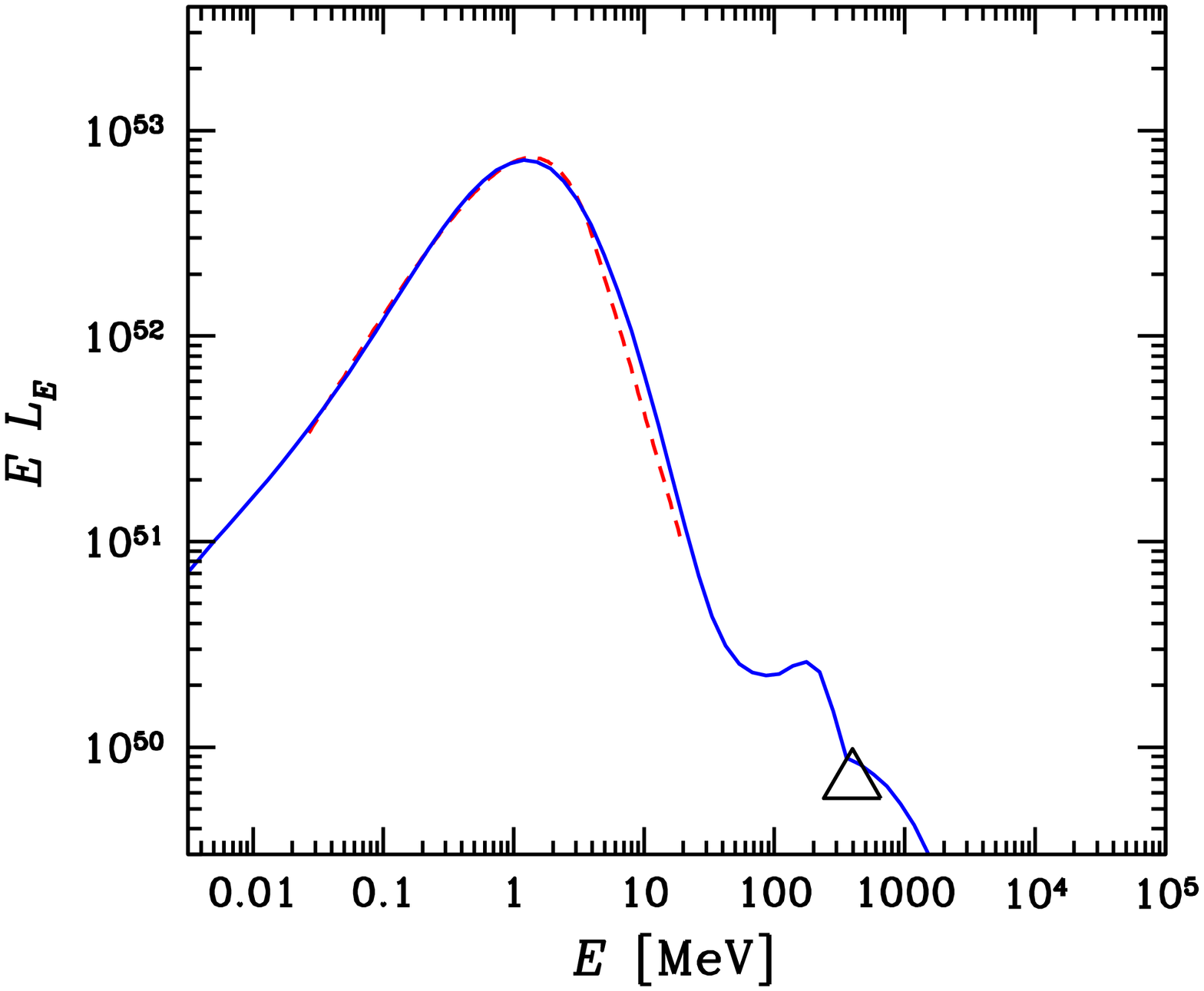}
\caption{Simulated spectrum of GRB 130427A (solid line), and a Band fit (dashed line, \citealt{Golenetskii13}).
Jet parameters: $\epsB=0.05$,
$\rcoll=3\times 10^{11}$~cm, $\Gamma(\rcoll)=64$, $n_{\rm ph}(\rcoll)/n_{\rm p}(\rcoll) = 1.2\times 10^4$,
$\Gamma_{\rm f} = 400$;
Heating parameters
$\ethinit+\enthinit=0.18$,
$\ethinit/\enthinit=1.4$, $\kth=\knth=0.04$;
the non-thermal dissipation law changes into $\knth=-2$ at $\taup=30$, the heating proceeds until $\tauT=2$.
The black triangle corresponds to the average flux above $100$~MeV observed by {\it Fermi}/LAT during the most intense phase of the prompt emission \citep{Ackermann14}.
}
\label{Fig:130427A}
\end{figure}

The results are shown in Figures~\ref{Fig:990123}-\ref{Fig:130427A}. The previous phenomenological fits 
are shown by red dashed curves
(see Table~\ref{tab:phenom} for fit parameters)
and our transfer model is shown by solid blue curves.
The achieved agreement demonstrates that the model is consistent with the data
and provides estimates for the jet parameters for each GRB.
The parameters are given in Table~\ref{tab:params}.

%%%%%%%%%%%%%%%%%%%%%%%%
%%%%%%%%%%%%%%%%%%%%%%%%

%%%%%%%%%%%%%%%%%%%%%%%%
%%%%%%%%%%%%%%%%%%%%%%%%

%%%%%%%%%%%%%%%%%%%%%%%%
%%%%%%%%%%%%%%%%%%%%%%%%

%%%%%%%%%%%%%%%%%%%%%%%%%%%%%%%%%%%
% BEGIN TABLE
\begin{table*}[t]
\begin{center}
\caption{
Jet parameters for the fitted GRBs.
\label{tab:params}}
\vspace{1.0mm}
\scriptsize{
\begin{tabular}{c|cccccccccc} \hline\hline
 GRB      & $L$\tablenotemark{a}			& $\Gamma(\rcoll)$\tablenotemark{b} 	& $\Gamma_{\rm f}$\tablenotemark{c}	& $\eta$\tablenotemark{d}	& $\epsilon_B$\tablenotemark{e} & $\epshinit$\tablenotemark{f} 	& $\ethinit/\enthinit$\tablenotemark{g} & $\kth$\tablenotemark{h}	& $\knth$\tablenotemark{i}   			& $\rph$\tablenotemark{j}	\\
          & $[10^{54} \ \mathrm{erg \,\, s}^{-1}]$  	& 		   			& 					&		 		&		 		&	 	  		& 	  	  	   		&  				&						& $[10^{13} \ \mathrm{cm}]$	\\  
          & 						& 		   			&					&		 		&		 		& 		  		&  		  	   		&  				&						&				\vspace{-0.3cm}	\\  \hline
 990123   & $0.44$ 					& $35$ 		   			& $500$					& $730$	 	 		& $0.018$	 		& $0.12$ 	  		& $4.7$  	  	   		& $-0.19$\tablenotemark{k}   	& $-0.19$\tablenotemark{k}			& $1$				\\
 090902B  & $2.3$ 					& $70$ 		   			& $1220$				& $1740$	 		& $0.012$	 		& $0.12$ 	  		& $0.85$  	  	   		& $-0.25$\tablenotemark{l}   	& $-0.25$\tablenotemark{l}			& $1$				\\
 130427A  & $0.43$ 					& $64$ 	   				& $400$					& $506$	 	 		& $0.050$	 		& $0.18$ 	  		& $1.4$ 	  	   		& $0.04$\tablenotemark{m}   	& $0.04\rightarrow -2$\tablenotemark{m}		& $2$				\\
 \hline
\end{tabular}
\tablenotetext{1}{Total jet luminosity (isotropic equivalent).}
\tablenotetext{2}{Jet Lorentz factor at $\rcoll$.}
\tablenotetext{3}{Final Lorentz factor $\Gamma_{\rm f}$ achieved by the jet at large radii
(calculated self-consistently from the model).}
\tablenotetext{4}{Jet energy per unit rest mass $\eta$.}
\tablenotetext{5}{Magnetization $\epsB$.}
\tablenotetext{6}{Total heating rate $\epshinit = \ethinit + \enthinit$.}
\tablenotetext{7}{Ratio of thermal to nonthermal heating $\ethinit/\enthinit$.}
\tablenotetext{8}{Radial scaling index of thermal dissipation $\kth$.}
\tablenotetext{9}{Radial scaling index of nonthermal dissipation $\knth$.}
\tablenotetext{10}{Photospheric radius $\rph$ (not a free parameter, calculated self-consistently from the model).}
\tablenotetext{11}{Dissipation proceeds until $\tauT=0.01$.}
\tablenotetext{12}{Dissipation proceeds until $\tauT=0.03$.}
\tablenotetext{13}{The nonthermal dissipation law changes at $\taup=30$. Dissipation proceeds until $\tauT=2$.}
}
\end{center}
\end{table*}
% END TABLE
%%%%%%%%%%%%%%%%%%%%%%%%%%%%%%%%%%%

%%%%%%%%%%%%%%%%%%%%%%%%%%%%%%%%%%%
% BEGIN TABLE
\begin{table*}[t]
\begin{center}
\caption{
Parameters of the phenomenological fits from the literature.
\label{tab:phenom}}
\vspace{1.0mm}
\scriptsize{
\begin{tabular}{c|ccccccc} \hline\hline
 GRB      	& $E_{\rm pk, obs}$\tablenotemark{a}	& $\alpha$\tablenotemark{b}	& $\beta$\tablenotemark{c}	& $\Gamma_{\rm pl}$\tablenotemark{d}	& $T-T_0$\tablenotemark{e}	& Instrument		& Reference 	  					\\
		& $[\mathrm{keV}]$  			& 		   			& 			&			 		& $[\mathrm{s}]$		&			&	 	  					\\  
		& 					& 		   			&			&		 	 		&				&			& 		  			\vspace{-0.3cm}	\\  \hline
 990123   	& $720$ 				& $-0.6$ 	   			& $-3.11$		& -	 	 	 		& $12-45$			& {\it CGRO}/BATSE	& \citet{Briggs99}   					\\
 090902B  	& $908$ 				& $0.07$ 	   			& $-3.9$		& $-1.94$	 	 		& $4.6-9.6$			& {\it Fermi}		& \citet{Abdo_090902B_l}, bin b				\\
 130427A  	& $1028$ 				& $-0.958$ 	   			& $-4.17$ 		& -	 		 		& $0-19$			& Konus/{\it Wind}	& \citet{Golenetskii13} 				\\
 \hline
\end{tabular}
\tablenotetext{1}{Spectral peak position $E_{\rm pk, obs}$ in the observer frame.}
\tablenotetext{2}{Low-energy slope $\alpha$.}
\tablenotetext{3}{High-energy slope $\beta$.}
\tablenotetext{4}{Power-law index in Band+PL fit.}
\tablenotetext{5}{Time interval of the fitted data relative to the trigger.}
}
\end{center}
\end{table*}
% END TABLE
%%%%%%%%%%%%%%%%%%%%%%%%%%%%%%%%%%%

The transfer model also provides a physical interpretation for the previously suggested
phenomenological components. In particular, the Band component below $\sim 10$~MeV
in all three bursts results from thermal Comptonization of low-energy photons by the heated 
plasma below the photosphere. 
The high-energy component from nonthermal Comptonization overlaps with the thermal 
Comptonization component and smoothly extends it beyond 10~MeV.
The associated spectral hardening is more gradual if
the thermal heating persists and dominates the dissipation
above the photosphere, as in e.g. GRB~990123.

The case of GRB~090902B is particularly interesting, as it shows strong 
deviations from the Band function at both low and high energies. These deviations 
were previously modeled as an additional power law component that extends from
the keV band up to GeV energies. Our transfer simulations show that the soft and hard 
excesses are produced by different radiative processes: synchrotron at low energies and 
IC at high energies. However, both are emitted by the same nonthermal
$e^\pm$ population, a result of strong nonthermal heating of the jet around the photosphere. 
The simplest mechanism of nonthermal heating is the decay of pions produced by
inelastic nuclear collisions (B10), which must operate in GRB jets 
unless they are strongly dominated by magnetic fields.
We find that GRB~090902B has $\epsB\approx 0.012$. This magnetization 
explains both the soft and hard excess. 
Remarkably, the same value of $\epsB$ is required by the shape of the
Band component observed in this burst.

The distinct spectral shapes in the three cases are mostly the result of different heating histories in the jet,
as well as different partitioning of the dissipated energy between thermal and non-thermal channels.
The absence of an extra component in GRB 990123 and GRB 130427A implies that non-thermal heating is weak in the
optically thin parts of the jet. Furthermore, the steep, almost cutoff-like appearance of the spectrum above the peak
in GRB 130427A suggests that most of the dissipation (thermal and nonthermal)
is confined to regions below the photosphere,
as heating at larger radii would tend to flatten the high-energy spectrum.
This last point is quite general
and can be applied to other bursts whose spectra have a cutoff power-law shape.
In contrast, the relatively soft low-energy slope of GRB 130427A
$\alpha\approx -1$
indicates that both dissipation channels have to be efficient at 
optical depths
$\tauT\gg 10$ below the photosphere, where the spectral slope 
$\alpha$ is formed.
The softening of $\alpha$ is also helped by the strong magnetization
$\epsB\approx 0.05$, which increases the supply of soft synchrotron photons.

Note that all three spectra peak just above 1~MeV, as is typical for bright GRBs.
The transfer model naturally explains the peak position. It is regulated by the 
photon number in the Wien zone,
which is calculated self-consistently with no fine tuning.

%####################################################

\section{Discussion}

\label{sec:disc}

In this paper we studied the production 
of radiation by opaque, dissipative, relativistic jets.
We used radiative transfer simulations to study the main features of the emission
mechanism and to compare the theoretical spectra with observations.
We presented a detailed physical model that fits the observed
GRB spectra and thus allows us to measure the main parameters of the jet.

We used a somewhat simplified
description of energy dissipation in the jet, with continual electron heating of two types, thermal 
and nonthermal. It is mainly motivated by the minimal heating model of B10
where electron/positron plasma receives energy from hot baryons through collisional 
processes, which are well defined and can be calculated from first principles. 
However, our simple parametrization of electron heating (power law scaling with radius, 
with a given slope and normalization) may also accommodate other heating models.

Another simplification adopted in this paper is the steady-state approximation
to radiative transfer (see B11 for a discussion of this approximation).
The theoretical spectra calculated
in this work represent the average over the causal 
timescale near the photosphere  $\delta t_\star\sim \rph/\Gamma^2c$, which is 
typically very short in the presented models,
comparable to a few ms.
The spectrum can evolve on timescales $\delta t>\delta t_\star$, as observed in GRBs.
We believe that this evolution is controlled by the noisy process
of jet formation and mass loading 
near the central engine, which is difficult to predict.

The main advantage of our transfer simulations is that they carefully include
relevant radiative processes, so that the radiation
spectrum is accurately calculated. 
Another advantage is that our dissipative jet model has a moderate number of parameters
(seven for an unbroken dissipation profile and nine if a break is required by the data).
Therefore, the transfer model can be used to fit observations. Unlike fits by 
phenomenological functions, fits by the physical model directly provide estimates for 
the main physical parameters of the GRB jets.

We find that
a continuously heated and moderately magnetized jet naturally produces a Band-type spectrum,
with spectral slopes $\alpha$, $\beta$, and peak position $\Epk$
consistent with observations.
The bulk of the observed photons in a typical GRB do 
not originate from the central engine or the vicinity of the Thomson photosphere;
instead they are produced 
in the opaque jet at optical depths $\tauT>100$.
The broadening of the spectrum into the final non-thermal shape takes place between
$\tauT\sim 10$ and $\sim 0.1$.

A typical burst with $L\approx 10^{52}$ erg/s, a canonical Band spectrum, and 
$\Epk\simlt 1$~MeV is reproduced by the model if: 
\begin{enumerate}
\item 
The jet magnetization is in the range $10^{-3}\lesssim\epsB\lesssim 0.1$.
Very weak magnetization increases $\Epk$ by suppressing synchrotron emission;
strong (near equipartition) magnetization softens the spectrum both below and above the peak,
and generates a prominent soft ``excess'' below a few tens of keV.
\item The jet Lorentz factor $\Gamma(\rcoll)\lesssim 100$
at radii comparable to that of the stellar progenitor.
Low $\Epk$ bursts are more ``photon-rich'' and require considerably lower $\Gamma(\rcoll)$ for more efficient photon production.
\item Heating has a nonthermal component that injects relativistic leptons into the jet.
The absence of non-thermal leptons would lead to hard spectra with high $\Epk$,
due to the lack of synchrotron emission.
\end{enumerate}

Our simulations demonstrate that
nonthermal heating in the sub-photospheric region 
loads the jet with $\epm$ pairs via pair-photon cascades.
In weakly magnetized jets the pairs outnumber protons by
a factor comparable to 10,
leading to the increase of the 
average photospheric radius by a similar factor.
The jets remain forever dominated by $e^\pm$ pairs, which will affect 
afterglow emission produced by the reverse shock when the jet is decelerated by
the external medium. The reverse shock emission might serve as a probe of 
pair loading.

\subsection{Spectral peak position}

Transfer simulations allow one to study the physical conditions in the jet,
how the prompt radiation spectrum is formed, what controls the observed spectral 
index, the position of the peak etc.

The evolution of the radiation spectrum in the expanding jet takes place in two stages
delineated by the Wien radius $\rW$ where Comptonization switches from saturated to 
unsaturated regime. 
The photon number accumulated in the spectral peak and its observed position 
$\Epk$ is determined near $\rW$.
Overall, we find that the dependence of $\Epk$ on parameters is 
rather weak and no fine-tuning is necessary to bring the peak into the observed range around 1~MeV.

Note that in all models presented in this paper we chose the conservative 
assumption that the jet is initially photon starved --- we chose a low initial 
photon number, which corresponds to a high initial $\Epk\approx 10$~MeV.
Then the exact initial condition is not important --- it is quickly ``forgotten'' as many 
more photons are generated at larger radii. This assumption becomes inconsistent 
for relatively slow jets, which have a large Planck radius (B13, V13) or for fast
jets with weak dissipation below the Wien radius. In these cases strong thermal radiation 
should be assumed at the inner boundary of the simulation.

The most efficient photon production mechanism,
which controls the observed $\Epk$ in our models,
is synchrotron emission
in the Wien zone $r<\rW$, well inside the photopsheric radius $\Rph$.
This mechanism works when a non-negligible fraction of the dissipated energy 
is channeled to nonthermal leptons. 
The constraints on jet magnetization in the photon production zone
are less restrictive than 
suggested in V13 and \citet{ThompsonGill14}.
We find that 
a moderate magnetization $\epsB\gtrsim 10^{-3}$
is sufficient to generate the photon number observed in a typical GRB.
V13 required $\epsB\sim 1$ because they neglected pair cascades,
which increase the number of synchrotron emitters, partially offsetting the 
reduction in synchrotron emissivity at small $\epsB$.
Also, the extended heating range considered in this work allows
more time to accumulate photons in the Wien peak.

\subsection{Low energy slope $\alpha$}

The typical photon index
of observed GRB spectra at $E<\Epk$ is $\alpha\sim -1$ \citep{Kaneko06}.
In the absence of synchrotron photon production, photospheric emission 
has a much harder spectrum
(\citealt{Peer2006,Giannios06}; B10; V11),
unless the jet has a very small collimation angle, comparable to $\Gamma^{-1}$
\citep{Lundman13}.

Our transfer simulations of magnetized jets with $\epsB>10^{-3}$ naturally 
explain the observed $\alpha\sim -1$ and its variations (Figure~\ref{fig:spectra:eB}).
The softening of the low-energy slope begins near the Wien radius where
the generated low-energy synchrotron photons are no longer Comptonized to
the spectral peak, ending up at intermediate energies.
For a broad range of parameters the resulting low-energy spectrum is roughly flat in
number of photons per $\ln E$ (i.e. $\alpha=-1$).
Similar low-energy slopes were recently obtained by \citet{ThompsonGill14}
in magnetically dominated jets. Despite the different regime,
the spectrum formation below the peak is similar:
unsaturated Comptonization of a low-energy photon source.

\subsection{Baryonic vs. Poynting-dominated jets}

In this work we have only considered moderately magnetized jets. 
Very weak magnetization $\epsB<10^{-3}$ is disfavored
based on the need for efficient photon generation and 
the observed spectral shape (in particular below the peak).
On the other hand, our results do not exclude the possibility of
a Poynting-dominated jet at small radii, as the initially dominant magnetic energy could have been dissipated at $r\ll\Rph$.

Radiation from magnetically dominated jets was recently considered by
\citet{ThompsonGill14}, \citet{GillThompson14} and \citet{BeguePeer15}.
Gill \& Thompson envision a two-stage evolution of the jet beginning from a baryon-free 
Poynting flux followed by baryon loading and photospheric emission.
In their picture,
the opacity is due to electron-positron pairs generated by dissipation,
which takes place in two separate episodes.
The $e^\pm$ pairs also generate photons through cyclo-synchrotron emission.
The evolution of radiative processes in their
scenario resembles that in our jet models: the generation of synchrotron photons 
is followed by Comptonization into a Band-like spectrum.

\subsection{Models for individual bursts}

We applied our radiative transfer model to three well-studied
bright bursts, GRB 990123, GRB 090902B, and GRB 130427A,
which show different prompt spectra.
Successful fits have been found in all three cases, giving estimates for the
main parameters $L$, $\eta$, $\epsB$, $\Gamma(\rcoll)$, $\epsth$, and $\epsnth$ (Table~1).
In particular, we find the jet magnetizations $\epsB=0.01-0.05$,
and the Lorentz factors between 400 and 1200.
The average photospheric radii $\Rph$ in the three bursts vary around $10^{13}$~cm.

These results suggest a systematic method for estimating the jet Lorentz factor 
in a larger sample of bursts, which is independent of another new 
method based on the reconstruction of the GeV+optical flash produced by the external 
blast wave at much larger radii $\simgt 10^{16}$~cm
\citep{BHV14,Vurm14,Hascoet15}.
For GRB 130427A we find $\Gamma_{\rm f} = 400$, which within uncertainties is
consistent with the value $\Gamma_{\rm ej} = 350$
obtained from the GeV+optical flash reconstruction \citep{Vurm14}. 
For GRB~090902B we find $\Gamma_{\rm f} \approx 1200$,
which is higher than $\Gamma_{\rm ej} = 600-900$ used to fit
the GeV flash \citep{Hascoet15}.
Note however that the reverse shock in GRB~090902B is relativistic; in this case 
there is a significant uncertainty in the upper limit on $\Gamma_{\rm ej}$
and the flash modeling only gives
a lower limit  $\Gamma_{\rm ej} > 600$ \citep{Hascoet15}.

In GRB~130427A we find that the nonthermal heating becomes
weak well before the jet expands to transparency.
Its nonthermal Band-like spectrum is
mainly the result of thermal Comptonization. The weak residual high-energy
emission from nonthermal heating still makes a significant contribution
to the GeV luminosity and is consistent with the 
{\it Fermi} LAT data during the main prompt emission episode.
It could also explain
the variability superimposed on the smooth extended GeV flash in GRB~130427A.

The featureless Band spectrum of GRB~990123 suggests that thermal heating dominates the dissipation also in this burst.
Our transfer model predicts a moderate excess below a few tens of keV due to synchrotron emission,
similar but weaker than that observed in GRB~090902B.
A hint of such excess is indeed seen in the data (Figure 2 in \citealt{Briggs99}).

In contrast, nonthermal dissipation in GRB~090902B is strong up to the 
photosphere and beyond. It well explains the observed high-energy component 
and the soft excess, which
were previously modeled as a power law of unknown origin \citep{Abdo_090902B_l}.
The high-energy component is also expected to make a contribution to the
GeV flash observed in GRB~090902B by {\it Fermi} LAT.
Our result for the Lorentz factor $\Gamma\approx 1200$ is similar to the 
estimate by \citet{Peer2012}. They used a different phenomenological model 
for GRB~090902B (a multicolor blackbody for the photosphere and nonthermal 
radiation from dissipation at a large radius) and estimated $\Gamma\approx 1000$.

Confining most of the dissipation to the subphotospheric region in bursts like GRB 130427A
is expected if dissipation is caused by neutron-proton collisions, whose rate 
declines at $\taup\lesssim 20$ (B10). The heating of thermal electrons by Coulomb collisions
with hot protons (stirred by n-p collisions) is also reduced at $\taupm \ll \mprot/(\me \ln{\Lambda})$,
where $\ln{\Lambda}\sim 20$ is the Coulomb logarithm (\citealt{Rossi2006}; B10).
Significant heating extending beyond the photosphere in GRB~090902B
suggests the presence of a different dissipation mechanism. For instance, internal shocks 
can occur both below and above the photosphere, and
the shocks can directly heat the photons and $\epm$ plasma  
without relying on n-p or Coulomb collisions \citep{Beloborodov16}.

Note that a systematic study of the entire parameter space was not attempted in this paper and is a currently ongoing work. 
Thus some degeneracies may be present between the parameters reported in Table~\ref{tab:params},
particularly if the actual dissipation profile is more complex than the ``minimal'' model used in this paper.
The relatively featureless spectra of GRB~990123 and GRB~130427A
are more prone to such degeneracies.
In contrast, the prominent extra component(s) in GRB~090902B make
this case more restrictive, thus the obtained solution and jet parameters
are most likely unique.

\subsection{Future prospects}

Future analysis of GRB spectra using transfer simulations can be developed
in two ways.
(1) The transfer models give spectra that could be observed with a high temporal 
resolution $\delta t_\star\sim \Rph/\Gamma^2c$, comparable to a few ms. 
The {\it evolving} nonthermal spectrum could be fitted by the model;
it would show the evolution of the jet parameters during the burst.
(2) Photon statistics in observed GRB spectra are usually accumulated on timescales 
$\delta t\gg \delta t_\star$, hindering the resolution of the instantaneous spectrum 
emitted by the variable jet. Then even the best time-resolved data analysis may give 
a mixture of different instantaneous spectra,
which could result in the presence of multiple components in the 
measured spectrum.
Transfer simulations may be used for the analysis of the 
multiple components.
A dominant Band component is almost always found in GRBs,
including the recent fits by the evolving mixture of three components:
Band + power law + thermal \citep{Guiriec15}.
The nonthermal emission may be well explained
by our model of a heated jet, including the excess at low and high energies
which was previously viewed as a separate 
power law (see Figure~\ref{Fig:090902B}).
The model also predicts
that emission from weakly heated jets has a quasi-thermal shape.
Thus the presence of a quasi-thermal component
\citep{Ryde04,Ryde05,RydePeer09,Guiriec15}
may indicate the presence of unresolved parts of the jet with weak heating.

The transfer model makes specific predictions for
polarization of the prompt radiation
(Lundman et al. 2016, in preparation), which can be tested by future observations.
The polarized radiation arises from synchrotron 
emission by nonthermal electrons;
it increases toward lower frequencies below the MeV peak and is strongest in the X-ray band.
Bursts with strong nonthermal dissipation extending to the photosphere 
(such as GRB~090902B) will be most promising for the detection of polarization.

\vspace{0.2in}
We thank Christoffer Lundman for helpful comments and discussions that helped to improve this manuscript.
This work was supported by NSF grant AST-1412485 and NASA ATP grant NNX15AE26G.

\appendix

\section{Dissipative jet dynamics}

\label{sec:app:dyn}

Let us rewrite the radiative transfer equation (\ref{eq:RTE}) in the form
\begin{align}
\frac{1}{r^2 \Gamma^2}\frac{\partial}{\partial\ln{r}} \left[ (1+\mu) r^2 \Gamma^2 \Inu \right] &= 
\frac{r}{\Gamma} (\jnu - \knu\Inu)	
 +(1+\mu)(1-g) \, \Inu 				\nonumber \\
&+ (1+\mu)(1 - g\mu) \frac{\d\Inu}{\d\ln\nu}
- \frac{\d}{\d\mu} \left[ (1-\mu^2)(1+\mu) \, g\Inu
\right].
\label{eq:RTE2}
\end{align}
Integrating Equation (\ref{eq:RTE2}) over the photon energy and taking the first two angular moments yields
\begin{align}
&\frac{1}{r^2 \Gamma^2}\frac{d}{d\ln{r}} \left[ r^2 \Gamma^2 (I_0 + I_1) \right] =
\frac{r}{\Gamma} (\jav - \kappa I_0) - g(I_0 - I_2), 						\label{eq:RTE0mom} \\
&\frac{1}{r^2 \Gamma^2}\frac{d}{d\ln{r}} \left[ r^2 \Gamma^2 (I_1 + I_2) \right] =
-\frac{r}{\Gamma} \, \kappa I_1 + g(I_0 - I_2),							\label{eq:RTE1mom}
\end{align}
where $\jav$ is the angle-averaged bolometric emissivity and we have assumed an isotropic (comoving) opacity $\kappa$.
The moments of intensity $I_1$, $I_2$ and $I_3$ are defined as
\begin{align}
I_{m} = \frac{1}{2}\int d\nu \int_{-1}^{1} I_{\nu}(\mu) \, \mu^m \, d\mu.
\label{app:eq:Imom}
\end{align}
The lab frame luminosity is given by $\Lrad = (4\pi)^2 r^2 \Gamma^2 (I_0 + 2I_1 + I_2)$
(using $\Gamma\gg 1$, see B11).
The equation governing its evolution is obtained by adding Equations (\ref{eq:RTE0mom}) and (\ref{eq:RTE1mom}),
\begin{align}
\frac{d\Lrad}{d\ln{r}} = (4\pi)^2 r^3 \Gamma \left[
\jav - \kappa (I_0 + I_1)
\right].
\label{eq:Lradevol2}
\end{align}
The dissipation rate in the comoving frame is $dE_{\rm h}/(dV^{\prime} dt^{\prime}) = 4\pi (\jav - \kappa I_0)$.
Transformation to the lab frame yields
\begin{align}
\frac{d\Lh}{d\ln{r}} = 4\pi r^3\Gamma \frac{dE_{\rm h}}{dV^{\prime} dt^{\prime}} = (4\pi)^2 r^3\Gamma \, (\jav - \kappa I_0).
\end{align}
Equation (\ref{eq:Lradevol2}) thus becomes
\begin{align}
\frac{d\Lrad}{d\ln{r}} = \frac{d\Lh}{d\ln{r}} - (4\pi)^2 r^3 \Gamma \, \kappa I_1.
\label{eq:Lradevol3}
\end{align}
The second term on the right hand side describes the work done by the radiation field
on accelerating the jet.

If dissipation is described as redistribution of the total energy 
$L=\Lpl+\Lrad=\mbox{const}$ from the bulk kinetic ($\Lpl$) to internal  (radiation-dominated)
form, one can write $\Lrad = L(1-\Gamma/\eta)$ to obtain a dynamical equation for the jet Lorentz factor
\begin{align}
\frac{d\Gamma}{d r} = -\eta \frac{d\epsh}{d r} + \sigmaT \Zpm \, \frac{4\pi I_1}{m_p c^3},
\label{eq:app:Gevol}
\end{align}
where $d\epsh/d\ln{r} = L^{-1} d\Lh/d\ln{r}$.
Alternatively,
the heat source can be modelled as a ``reservoir'' of free energy carried by the jet 
in the form of internal bulk motions, $\Lturb$. Then $L=\Lpl+\Lrad+\Lturb=\mbox{const}$, and
the energy balance is given by
\begin{align}
\frac{d\Lh}{d\ln{r}} = \frac{d\Lrad}{d\ln{r}} + \dotM c^2 \frac{d\Gamma}{d\ln{r}}.
\end{align}
In this case, the dynamical equation becomes
\begin{align}
\frac{d\Gamma}{d r} = \sigmaT \Zpm \, \frac{4\pi I_1}{m_p c^3}.
\label{eq:app:Gevol2}
\end{align}

In the optically thick domain it is useful to rewrite the dynamical equation in a form where the (small) first moment $I_1$ of the intensity does not appear.
In such case $I_2 \approx I_0/3 \gg I_1$
and $\Lrad = (4/3)(4\pi)^2 r^2 \Gamma^2 I_0$, whereby Equation (\ref{eq:RTE0mom}) can be written as
\begin{align}
\frac{d\Lrad}{d\ln{r}} = \frac{4}{3} \frac{d\Lh}{d\ln{r}} - \frac{2g}{3} \Lrad,
\label{eq:app:Levol}
\end{align}
where the last term accounts for adiabatic cooling.
In place of Equation (\ref{eq:app:Gevol}) one now obtains
\begin{align}
\frac{d\ln{\Gamma}}{d\ln{r}} = \frac{1}{2 + \Gamma/\eta} \left[
2\left( 1 - \frac{\Gamma}{\eta} \right) - 4\frac{d\epsh}{d\ln{r}}
\right].
\label{eq:app:Gevol_th}
\end{align}
In the absence of dissipation Equation (\ref{eq:app:Gevol_th}) can be integrated straightforwardly to yield the standard
solution for $\Gamma$ in baryonic (initially) radiation dominated jets
\citep[see e.g.][]{PSN93}.

In the reservoir model of the heat source
Equation (\ref{eq:app:Gevol2}) is replaced by
\begin{align}
\frac{d\ln{\Gamma}}{d\ln{r}} =
\frac{2\Lrad - d\Lh/d\ln r}{3\Gamma\dotM c^2 + 2\Lrad}.
\label{eq:app:Gevol2_th}
\end{align}

\section{Pair-loading in a dissipative jet}

\label{sec:app:pload}

The cooling of the injected $\epm$ pairs
and their subsequent thermalization is very efficient
in the high compactness environment of GRB jets.
Thus bulk of the $\epm$ pairs reside in a Maxwellian distribution at any given time.
Their density is controlled by injection of new pairs (both primary particles as well as secondaries from the cascade),
$\epm$ annihilation, and expansion of the jet (cf. B10),
\begin{align}
\frac{c}{r^2}\frac{d}{dr}\left(
\Gamma r^2 \npm \right)
= \ninj - \nann,
\label{eq:app:pbal}
\end{align}
where
\begin{align}
\ninj = \frac{Y\epsinj L}{4\pi\me c^2 r^3 \Gamma}, \qquad
\qquad
\nann =\frac{3}{4} c\sigmaT \, n_+ n_-
=\frac{3}{16} c\sigmaT \, (\npm^2 - \nprot^2),
\end{align}
and $\npm = n_+ + n_- = 2 n_+ + \nprot$.
The pair yield $Y$ characterizes the fraction of energy injected into the primary pairs
that ends up in pair rest mass through the pair-photon cascade.
Defining the pair loading $\Zpm = \npm/\nprot$, and using $\Gamma r^2 \nprot = \dotM/(4\pi \mprot c)=\mbox{constant}$,
Equation (\ref{eq:app:pbal}) can be written as
\begin{align}
\frac{d\Zpm}{d\ln{r}} =
\frac{Y\epsinj L}{\Gamma\dotM c^2} \frac{\mprot}{\me} - \frac{3}{16} \taup (\Zpm^2 -1),
\label{eq:app:pbal2}
\end{align}
where $\taup = \sigmaT\nprot r/\Gamma$.

Two regimes can be identified in Equation (\ref{eq:app:pbal2}): (1) creation-annihilation balance, where both terms on the RHS are larger than the LHS,
and (2) freezeout, where annihilation (last term) is negligible.
In creation-annihilation balance the pair loading evolves as
\begin{align}
\Zpm = \left(
\frac{16 Y\epsinj L}{3\taup\, \Gamma\dotM c^2} \frac{\mprot}{\me} +1
\right)^{1/2}.
\end{align}
Annihilation freezes out once $\taupm = \Zpm\taup = 16/3$;
beyond this radius the pair loading
is governed by the first two terms in Equation (\ref{eq:app:pbal2}) and
can increase at most by a logarithmic factor in radius
(assuming a logarithmically flat heating rate and $\Gamma=\mbox{const}$).

\section{Photon production by synchrotron emission}

\label{sec:app:phprod}

The rate at which photons are generated and 
accumulated in the spectral peak
depends on the interplay between synchrotron emission,
reabsorption, IC scattering and induced Compton downscattering.
It was shown in V13 that
competition between these processes defines
a characteristic photon energy above which the emitted synchrotron photons
are upscattered to the Wien peak rather than 
downscattered and/or reabsorbed.
Below we extend the results of V13 to include pair-photon cascades
and apply them to the extended heated Wien zone. 

At high optical depths the timescales for all relevant radiative processes
are much shorter than the jet expansion timescale.
In this regime the Kompaneets equation can be written as
\begin{align}
-\frac{\sigma_T \Ne}{m_ec} \frac 1{E^2}\frac{\partial}{\partial E}E^4\left(\kB \Te\frac{\partial \nocc}{\partial E}+\nocc^2\right)
=\jocc(E)-c\, \kappa(E)\nocc,
\label{eq:meth:Komp}
\end{align}
where $\nocc$ is the photon occupation 
number, $\jocc$ is the emissivity (here we consider only synchrotron) in units of the number of photons per 
unit phase space volume per second,
$\kappa$ is the (synchrotron absorption) opacity,
$\Ne$ is the density of the thermal electron/positron component,
and $\Te$ is the electron temperature.
We have neglected the recoil term in Equation (\ref{eq:meth:Komp}) since
photon generation takes place at $E\ll\kB\Te\ll \me c^2$ where recoil losses are negligible.

By comparing the terms in Equation (\ref{eq:meth:Komp})
one can identify three regimes where different physical processes dominate.
At low enough photon energies the radiation has the usual
optically thick synchrotron spectrum for power-law electrons,\footnote{Here 
     we are using the delta-function approximation, assuming that all synchrotron 
     photons from a single electron are emitted at $E=0.3\gamma^2\EB$.}
\begin{align}
\nocc=\frac{\jocc}{c\kappa} = \frac{1}{p+2} \left( \frac{E}{\EB} \right)^{1/2}
\left( \frac{E}{\me c^2} \right)^{-1}.
\label{eq:ssa}
\end{align}
As the synchrotron emitting electrons are relativistic,
the brightness temperature of the optically thick synchrotron radiation 
$\TB\equiv E\nocc/\kB \gg \Te$.
In this case the induced scattering term dominates the Kompaneets operator,
tending to downscatter the synchrotron photons.
 The first break in the spectrum occurs where the induced scattering rate 
becomes dominant over the reabsorption rate.
Above this energy (but before the second break, see below) the Kompaneets equation reduces to
\begin{align}
-\frac{\sigma_T \Ne}{m_ec} \frac 1{E^2}\frac{\partial}{\partial E}E^4 \nocc^2
=\jocc(E).
\label{eq:meth:Ind}
\end{align}
The synchrotron emissivity of a power-law distribution of electrons, 
$\Ne(\gamma)=n_0 \gamma^{-p}$,
is
\begin{align}
\jocc(E) = j_0 \left( \frac{E}{\EB} \right)^{-(p+5)/2} = \frac{\pi}{4} \afs^{-1} c\sigmaT n_0
\left( \frac{\me c^2}{\EB} \right)^2 \left( \frac{E}{\EB} \right)^{-(p+5)/2},
\label{jsyn}
\end{align}
where $\afs=e^2/\hbar c$ if the fine structure constant and 
$\EB=\hbar\, eB/m_ec$.
The solution of Equation (\ref{eq:meth:Ind}) takes the form
\begin{align}
\nocc=\left[\frac{2m_ec \, \jocc(E)}{(p-1)\sigma_{\rm T}\Ne E}\right]^{1/2} \propto E^{-(p+7)/4}.
\label{ind_compt}
\end{align}

Compton upscattering begins where
the radiation brightness temperature decreases down to the electron temperature.
This yields the second break in the spectrum, above which the usual thermal Comptonization spectrum is established.
The characteristic energy, $E_0$, can be found by equating the two terms in the Comptonization operator
in Equation (\ref{eq:meth:Komp}),
\begin{align}
-\kB \Te\frac{\partial \nocc}{\partial E} = \nocc^2.
\end{align}
Using the spectrum (\ref{ind_compt}), this yields
\begin{align}
\frac{p+7}{4}\, \frac{\kB \Te}{E_0} = \nocc(E_0),
\label{nE_0}
\end{align}
or equivalently
\begin{align}
\frac{(p-1)(p+7)^2}{32}\, \frac{(\kB \Te)^2}{E_0} \, \frac{\sigmaT \Ne}{\me c} = \jocc(E_0).
\label{jE_0}
\end{align}

To proceed, one has to specify the electron distribution.
Let us assume that relativistic electrons are continuously injected
at a fixed Lorentz factor $\ginj$ with power $\Qinj$
$[\mathrm{erg} \cdot\mathrm{cm}^{-3}\cdot\mathrm{s}^{-1}]$.
The injected high-energy pairs initiate a pair-photon cascade with a large number of secondaries.
Let us denote their number per injected primary particle as $\Ms$.
Denoting the pair yield as $Y=\Ms/\ginj$
we can express the equilibrium pair distribution in the cascade as
\begin{align}
\Ne(\gamma) = n_0 \gamma^{-p},
\end{align}
where
\begin{align}
n_0 = \frac{3Y\gcascmin^{\alphacasc}}{4c\sigmat} \frac{\epsinj}{(\epsB + \epsrad)} \frac{c\Gamma}{r},
\label{eq:n0:casc}
\end{align}
$p=2+\alphacasc$, $\gcascmin$ is the electron energy below which the saturated cascade turns off, and
$\alphacasc\approx\ln{\Ms}/(\ln{\ginj} - \ln{\gcascmin})$.
The energy fractions $\epsB$ and $\epsrad$ are defined via
\begin{align}
\UB = \frac{\epsB  L}{4\pi c r^2 \Gamma^2} \quad \mbox{and} \quad \Urad = \frac{\epsrad  L}{4\pi c r^2 \Gamma^2},
\end{align}
and $\epsinj$ is the fraction of total available energy deposited into the injected electrons per logarithmic radius interval,
\begin{align}
\epsnth L = \frac{d \Lnth}{d\ln r}
= 4\pi r^3 \Gamma\Qinj.
\end{align}

Using $n_0$ given by Equation (\ref{eq:n0:casc}) in the expression (\ref{jsyn}) for the emissivity
and inserting the latter into (\ref{jE_0}) yields the critical photon energy/electron Lorentz factor
\begin{align}
  \left(\frac{E_0}{\EB}\right)^{(p+3)/2} = (\sqrt{0.3} \, \gamma_0)^{p+3}=
  \frac{96\pi}{(p-1)(p+7)^2} 
  \afs^{-1} \, \frac{\epsinj \, Y\gcascmin^{\alphacasc}}{(\epsB + \epsrad)} \, \frac{\tauT}{y^2} 
  \left(\frac{\EB}{\me c^2}\right)^{-1}.
\label{eq:E0:casc}
\end{align}
Here $\tauT = \sigmaT\Zpm\nprot r/\Gamma$, $y=4\tauT\thetae$ is the Compton parameter and
the baryon number density is given by
\begin{align}
\nprot =
\frac{L}{4\pi \mprot c^3 r^2 \Gamma \eta}.
\end{align}
The value of the critical Lorentz factor $\gamma_0$
is typically between a few and a few tens.
If the pair cascades are suppressed (e.g. by synchrotron cooling), one obtains
\begin{align}
\gamma_0 = 7.8 \, L_{52}^{-3/10} \, R_{10}^{2/5} \, \Gamma_1^{3/5} \, \eta_3^{1/5} \, \gamma_{{\rm inj}, 2}^{-1/5} \,
\theta_{-2}^{-2/5}  \, \epsB^{-1/10} \, \left( \frac{\epsinj}{\epsB + \epsrad} \right)^{1/5}.
\end{align}

The synchrotron photon production rate at $E>E_0$ is
\begin{align}
\dotNsynch =  \frac{4\pi}{(ch)^3} \int_{E_0} \jocc(E) E^2 \, dE =
\frac{3}{2(p-1)}  \frac{\epsinj Y\gcascmin^{\alphacasc}}{(\epsB + \epsrad)} \frac{c\Gamma}{r} \, \frac{\UB}{\EB} \left(\frac{E_0}{\EB}\right)^{-(p-1)/2}.
\label{eq:dotN:casc}
\end{align}
At a sphere of radius $r$, the number flux of photons
produced between $r_{\rm min}$ and $r$ is 
\begin{align}
\dotNph(r) = 
\int_{r_{\rm min}}^r 4\pi r^3 \, \dotNsynch(r) \, d\ln{r}.
\label{eq:dotNph}
\end{align}

The quantities $\epsrad(r)$ and $\Gamma(r)$
are determined by the heating model, in particular by $d\Lh/d\ln{r}$, see
Equations (\ref{eq:app:Levol}) and (\ref{eq:app:Gevol2_th}).
One last quantity entering the photon production rate is the electron temperature.
In the Wien zone it can be determined from the relation
\begin{align}
   \Lrad = \frac{4}{3} \epsrad L = 4\Gamma\kB \Te \,\dotNph.
\label{eq:app:Lwien}
\end{align}
For given cascade parameters $Y$ and $\gcascmin$, 
Equations (\ref{eq:E0:casc}) -- (\ref{eq:app:Lwien}) along with (\ref{eq:app:Levol}) and (\ref{eq:app:Gevol2_th}) form a closed set.
Its solution gives the number of synchrotron photons $\dotNph(r)$ 
which are Comptonized toward the Wien peak. 
The solution is simplified if one uses the approximation
$\dotNph \approx 4\pi r^3 \, \dotNsynch$.

It is convenient to express the result
as the number of produced photons per 
proton, $\nph/\nprot=\dotNph/\dot{N}_{\rm p}$, where $\dot{N}_{\rm p}=L/(\mprot c^2\eta)$ 
is the proton number flux carried by the flow. In the absence of $e^\pm$ cascade, 
$\alphacasc=0$ and $Y=\ginj^{-1}$, one finds
\begin{align}
    \frac{\nph}{n_p}
= 4.1 \times 10^6 \, L_{52}^{-1/7} \, 
     r_{10}^{3/7}
  \, \Gamma_1^{-5/7} \, 
\eta_3^{6/7} \, \gamma_{{\rm inj}, 2}^{-4/7} \,
\epsB^{3/7} \, \epsrad^{2/7} \, \left( \frac{\epsinj}{\epsB + \epsrad} \right)^{4/7}.
\label{eq:app:Nsynch}
\end{align}
In the opposite limit of a fully saturated cascade one obtains
\begin{align} 
   \frac{\nph}{n_p} = 1.83 \times 10^7 \, 
       r_{10}^{1/5}
   \, \Gamma_1^{-1} \, \Zpm^{1/5}
   \eta_3^{4/5} \epsB^{2/5} \, \epsrad^{2/5} \, \left( \frac{\epsinj}{\epsB + \epsrad} \right)^{2/5},
\label{eq:app:Nsynch2}
\end{align}
where we used $p=3$ ($\alphacasc=1$).
In a general case, the radial dependence of photon production is 
given by
\begin{align}
  \frac{\nph}{n_p}
  \propto r^{\frac{5-p}{3p+1}} \, \Gamma^{\frac{-5p+5}{3p+1}} \, \Zpm^{\frac{p-1}{3p+1}}
  \epsrad^{\frac{2p-2}{3p+1}}
  \left(\frac{\epsinj Y\gcascmin^{\alphacasc}}{\epsB + \epsrad} \right)^{\frac{4}{3p+1}}.
\label{eq:app:Nsynch3}
\end{align}

The above analysis neglects
some effects that are not straightforward to describe analytically.
In particular, near the critical Lorentz factor $\gamma_0\sim 10$,
the IC cooling is modified by the Klein-Nishina effect, 
as the target photon energy in the electron rest frame $\sim3\kB\Te \gamma_0$ is
comparable to $\me c^2$.
The Klein-Nishina recoil effect also influences the fate of the IC photons --- they can
scatter and lose energy to electron recoil before annihilating into pairs. This effect is 
particularly important for those IC photons that see a low opacity for pair production, which 
happens if they are below the threshold for interacting with the Wien-peak radiation. These
effects can substantially alter the number of electrons near $\gamma_0$,
and consequently the number of produced synchrotron photons.
The above analytic estimates can only serve as 
a rough guide to the photon production showing its trends, i.e. the
dependence on $r$, $\epsB$, $\Gamma$ etc.
The accurate photon number is provided by our numerical simulations.

\vspace{5mm}

% \bibliographystyle{apj} 
% \bibliography{biblio}

\end{document}